\documentclass[superscriptaddress,
                amsmath,
                amssymb,
                aps,
                longbibliography]{revtex4-2}

\usepackage{bm}             
\usepackage{hyperref}
\usepackage{graphicx}
\usepackage{multirow}
\usepackage{nicefrac}       
\usepackage{color}
\usepackage{mathtools}
\usepackage{placeins}
\def \vector#1{\mathbf{#1}}
\def \matrix#1{\mathbf{#1}}

\newcommand{\revise}{\textcolor{black}}
\begin{document}

\title{Data-driven computation of adjoint sensitivities without adjoint solvers: \\ An application to thermoacoustics}
\author{Defne E. Ozan}
\affiliation{Imperial College London, Department of Aeronautics, Exhibition Road, London SW7 2BX, UK}
\author{Luca Magri}
\affiliation{Imperial College London, Department of Aeronautics, Exhibition Road, London SW7 2BX, UK}
\affiliation{The Alan Turing Institute, London NW1 2DB, UK}
\affiliation{Politecnico di Torino, DIMEAS, Corso Duca degli Abruzzi, 24 10129 Torino, Italy}

\begin{abstract}
Adjoint methods have been the pillar of gradient-based optimization for decades. They enable the accurate computation of a gradient (sensitivity) of a quantity of interest with respect to all  system’s parameters in one calculation. When the gradient is embedded in an optimization routine, the quantity of interest can be optimized for the system to have the desired behaviour. Adjoint methods, however, require \revise{the system’s governing equations and their Jacobian}. In this paper, we propose a computational strategy to infer the adjoint sensitivities from data \revise{when the governing equations might be unknown (or partly unknown), and noise might be present.} The key component of this strategy is an echo state network, which learns the dynamics of nonlinear regimes with varying parameters, and it evolves dynamically via a hidden state. Although the framework does not make assumptions on the dynamical system, we focus on thermoacoustics, which are governed by nonlinear and time-delayed systems. First, we show that a parameter-aware Echo State Network (ESN) infers the parameterized dynamics. Second, we derive the adjoint of the ESN to compute \revise{two types of sensitivity: (i) parameter sensitivity, which is the gradient of a time-averaged cost functional with respect to physical/design parameters of the system; and (ii) initial condition sensitivity, which is the gradient of a cost functional of the final state with respect to the initial condition.} Third, we propose the Thermoacoustic Echo State Network (T-ESN), which \revise{embeds} the physical knowledge in the network architecture \revise{for improved generalization}. Fourth, we apply the framework to a variety of nonlinear thermoacoustic regimes of a prototypical system. We show that the T-ESN accurately infers the correct adjoint sensitivities of the acoustic energy with respect to the flame parameters \revise{and initial conditions}. The results are robust to noisy data, from periodic, through quasiperiodic, to chaotic regimes. The inferred adjoint sensitivities are employed to suppress an instability via steepest descent. We show that a single network predicts the nonlinear bifurcations on unseen scenarios, which allows it to converge to the minimum of the acoustic energy. This work opens new possibilities for gradient-based data-driven design optimization. 
\end{abstract}

\maketitle

\section{Introduction}
Engineers and scientists strive to design flow and acoustic systems that behave in an optimal way~\citep[e.g.,][]{peter2010NumericalSensitivityAnalysis,sipp2010DynamicsControlGlobal,luchini2014AdjointEquationsStability,camarri2015FlowControlDesign,magri2019AdjointMethodsDesign}. Finding the global optimum is challenging because the temporal dynamics can be nonlinear, time-delayed, and chaotic, which makes the objective function non-convex~\citep[e.g,][]{huhn2020StabilitySensitivityOptimisation,huhn2022GradientfreeOptimizationChaotic}. A practical approach is to solve the optimization problem by gradient-descent, in which the objective function is optimized along the gradient direction until a local optimum is found. The gradient (also known as sensitivity) offers the information about how the quantity of interest changes with the system's parameters. Central to gradient-based optimization methods are adjoint methods. Adjoint methods enable the accurate computation of the gradient with a computational cost that is not affected by the number of design parameters. For example, if the system has $N$ parameters, adjoint methods compute the gradient of a objective function with only one calculation. This \revise{is in} contrast with tangent linear or finite difference methods, which would require $N$ computations~\citep[e.g.,][]{giles2000IntroductionAdjointApproach,gunzburger2002PerspectivesFlowControl,giannetti2007StructuralSensitivityFirst,
sipp2010DynamicsControlGlobal,
luchini2014AdjointEquationsStability,
magri2019AdjointMethodsDesign}.

The optimization of a quantity of interest of a nonlinear system with adjoint methods requires, in general, six ingredients: 
(i) The system's nonlinear governing equations are integrated forward in time to obtain the baseline (direct) solution; 
(ii) the nonlinear equations are linearized to obtain the Jacobian of the system; 
(iii) the adjoint equations are derived from the Jacobian, 
(iv) the adjoint equations are initialized at the end of the direct solution, and are integrated backwards in time (because in nonlinear systems the Jacobian depends on the direct solution, strategies such as checkpointing become often necessary to decrease the memory requirements~\citep{griewank2000Algorithm799Revolve}); 
(v) at the end of the adjoint integration, the gradient of the objective function to all the design parameters of interest can be evaluated by an inner product; and 
(iv) the gradient is embedded in a gradient-based optimization scheme to iteratively compute the optimum of the objective function. This process is also known as the direct-adjoint looping. 
On the one hand, in finite difference  or tangent linear approaches, 
the computational cost increases linearly with the number of parameters because the sensitivity calculation needs to be performed for each parameter. On the other hand, the adjoint method computes the gradient of the objective with respect to all parameters with a single operation, which makes it computationally cheaper when the number of parameters is larger than the number of objective functionals~\citep[e.g.,][]{pironneau1974OptimumDesignFluid,jameson1988AerodynamicDesignControl, jameson1998OptimumAerodynamicDesign, jameson1999ReEngineeringDesignProcess,giles2000IntroductionAdjointApproach}. For this reason, adjoint methods have found extensive application in design optimization, and  both linear and nonlinear sensitivity analysis of fluid dynamics applications. 
To bypass the need of the system’s equations, a data-driven model can be employed. The key idea of this paper is to learn an approximation of the dynamical systems from observables, and then compute the adjoint of it. This approach is useful when working with experimental data, or data whose equations are not fully known.  
To infer the governing dynamics from observables (data), we employ a type of reservoir computer, the Echo State Network (ESN)~\citep{lukosevicius2012PracticalGuideApplying}, which is a universal approximator and a type of recurrent neural network (RNN)~\citep{grigoryeva2018EchoStateNetworks}. The ESN's training is  straightforward  because it does not require backpropagation. Its Jacobian can also be computed in a straightforward way due to the shallow architecture~\citep{margazoglou2023StabilityAnalysisChaotic}. ESNs have been shown to successfully (i) make time-accurate short-term forecasts~\cite{pathak2018ModelFreePredictionLarge,
doan2020PhysicsinformedEchoState,racca2021RobustOptimizationValidation}, (ii) reconstruct unmeasured variables~\citep{lu2017ReservoirObserversModelfree,doan2020LearningHiddenStates}, (iii) learn ergodic properties, such as Lyapunov exponents, and long-term statistics of chaotic flows~\cite{pathak2017UsingMachineLearning,huhn2022GradientfreeOptimizationChaotic,margazoglou2023StabilityAnalysisChaotic}, and (iv) infer model errors in thermoacoustics~\citep{novoa2024InferringUnknownUnknowns}. In recent advances, low-order representations of turbulence were learned from high-dimensional data, and the dynamics of this low-dimensional system was propagated with the ESN~\citep{racca2023PredictingTurbulentDynamicsa}. In this paper, we take an additional step. Because in order to perform design optimization of system's parameters, we require a parameter-aware extension of the Echo State Network~\citep{xiao2021PredictingAmplitudeDeath}, which learns the parameterized dynamics and can operate across parameter regimes. \revise{In a recent study, \citep{ozan2024AdjointSensitivitiesChaotic} demonstrated the proposed data-driven sensitivity framework on a prototypical chaotic system, and obtained the adjoint sensitivities of long-time averages to the system's parameters.}

\revise{In this paper, we develop a data-driven sensitivity framework, with an application on  nonlinear and time-delayed thermoacoustic oscillations.} These are phenomena that can be detrimental in the manufacturing of gas turbines~\citep{lieuwen2006CombustionInstabilitiesGas}, and rocket engines~\citep{culick2006UnsteadyMotionsCombustion}. In thermoacoustic systems, if the heat release and acoustic pressure are sufficiently in phase,  self-sustained oscillations can occur~\citep{rayleigh1878EXPLANATIONCERTAINACOUSTICAL,magri2020SensitivityRayleighCriterion}. Therefore, their prediction, control, and suppression are important both during design and during operation. This is especially challenging because thermoacoustics exhibit nonlinear behaviour and bifurcations to periodic, quasiperiodic, and chaotic regimes~\citep{kabiraj2012BifurcationsSelfExcitedDucted}, which motivates the development of methods for nonlinear analysis. Preliminary design of thermoacoustic systems is performed by linear stability~\citep{lieuwen2006CombustionInstabilitiesGas, juniper2018SensitivityNonlinearityThermoacoustic}. Linear stability analysis determines the growth rates of infinitesimal perturbations to the mean flow, i.e., the eigenvalues of the linearized system of equations around the mean flow~\citep{magri2023LinearFlowAnalysis}. 
To name a few examples, adjoint-based linear stability analysis enabled calculation of sensitivity to parameters and passive control mechanisms~\citep{magri2013SensitivityAnalysisTimedelayed}, optimal placement and tuning of acoustic dampers~\citep{mensah2017AcousticDamperPlacement} and stabilization of thermoacoustic oscillations by geometry changes~\citep{aguilar2018AdjointMethodsElimination}. 
An in-depth review of development of adjoint methods for thermoacoustics can be found in~\citep{magri2019AdjointMethodsDesign, magri2023LinearFlowAnalysis}. Beyond linear stability analysis, adjoint methods were used for weakly nonlinear analysis to predict subcritical Hopf bifurcations~\citep{orchini2016WeaklyNonlinearAnalysis}, and Floquet analysis to calculate the stability and the period of limit cycle oscillations~\citep{magri2019AdjointMethodsDesign}. The stability of chaotic thermoacoustics was studied by~\citep{huhn2020StabilitySensitivityOptimisation} using covariant Lyapunov analysis, who analytically showed that this is equivalent to eigenvalue and Floquet analysis for fixed point and limit cycle solutions, respectively. 
As a result,~\citep{huhn2020StabilitySensitivityOptimisation} achieved gradient-based minimization of acoustic energy in chaotic thermoacoustics using shadowing methods. The application of this method was challenged by the discontinuities observed in the acoustic energy when bifurcations occur, computational cost of shadowing methods, and non-hyperbolicity of thermoacoustics for some regimes. To circumvent these issues, a gradient-free optimization method based on Bayesian optimization was proposed~\citep{huhn2022GradientfreeOptimizationChaotic}. Among applications of data-driven methods in thermoacoustics, the ESN was previously used to predict the nonlinear dynamics~\citep{huhn2022GradientfreeOptimizationChaotic} and calculate the Lyapunov exponents~\citep{margazoglou2023DataDrivenStabilityAnalysisa} of a chaotic regime, and infer the model error of a low-order model~\citep{novoa2024InferringUnknownUnknowns}. 

The overarching objective of this paper is to address the challenge of developing code-specific adjoint solvers, which requires the system's equations and the Jacobian to be known. We propose data-driven nonlinear direct-adjoint looping to compute \revise{sensitivities to the system's parameters and initial conditions}, and apply this to a prototypical thermoacoustic system. The specific objectives are to \revise{introduce} a robust parameter-aware ESN architecture for this purpose, derive its adjoint, and demonstrate the data-driven inference of parameterized dynamics and associated adjoint sensitivities. We refer to this architecture as the thermoacoustic ESN and show that (i) it learns the parameterized dynamics of nonlinear, time-delayed thermoacoustics, (ii) predicts bifurcations, (iii) accurately obtains the adjoint sensitivities, (iv) the acoustic energy is minimized in a gradient-based optimization framework, and (v) demanding scenarios such as training with noisy or chaotic data can be handled.

This paper is organized as follows. In Sec.~\ref{sec:background}, we give a background on the mathematical framework for sensitivity analysis and the parameter-aware ESN. In Sec.~\ref{sec:adjoint_esn}, we derive the adjoint of the ESN. In Sec.~\ref{sec:thermoacoustics}, we present the nonlinear and time-delayed thermoacoustic model, namely the Rijke tube. In Sec.~\ref{sec:thermoacoustic_esn}, we introduce the time-delayed ESN designed for thermoacoustics and derive its adjoint. In Sec.~\ref{sec:results_learn}, we show the performance of the ESN in terms of prediction and accuracy of the learned adjoint sensitivities, and employ it for the gradient-based design optimization of thermoacoustics in Sec.~\ref{sec:results_grad}. We address training with noisy data in Sec.~\ref{sec:results_noise}, and learning chaotic regimes in Sec.~\ref{sec:results_chaos}. \revise{We show the sensitivity to initial conditions in Section~\ref{sec:init_sensitivity}. A discussion is offered in  Section~\ref{sec:limit}.} Section~\ref{sec:conclusion} concludes the paper.

\section{Background}\label{sec:background}
\subsection{Sensitivity Analysis}\label{sec:background_sensitivity}
We consider a dynamical system in the form of
\begin{equation}
    \dot{\vector{x}}-\vector{F}(\vector{x},\vector{p}) = 0, 
    \label{eq:cont_constraint}
\end{equation}
where $\vector{x} \in \mathbb{R}^{n_x}$ is the state vector, $\dot{\vector{x}}$ is its time derivative, $\vector{p} \in \mathbb{R}^{n_p}$ is the parameters' vector, and $\vector{F}: \mathbb{R}^{n_x} \times \mathbb{R}^{n_p} \rightarrow \mathbb{R}^{n_x}$ is the operator that describes the system's evolution. We assume that $\vector{F}$ is smooth so that the Jacobian of the system, i.e., $\partial\vector{F}/\partial\vector{x}$, is defined everywhere, which is a requirement for gradient-based sensitivity analysis. 
We wish to optimize an objective functional, $\mathcal{J}$, which is a time-average of $\tilde{\mathcal{J}}$~\cite{magri2019AdjointMethodsDesign}
\begin{equation}
    \mathcal{J} = \frac{1}{T}\int_{0}^{T} \tilde{\mathcal{J}}(\vector{x}(t), \vector{p})dt,
    \label{eq:cont_objective}
\end{equation}
subjected to the system dynamics~\eqref{eq:cont_constraint}. A straightforward approach to obtain sensitivity information is to apply a small perturbation, $h$, to the parameters of interest, $p_{i} \in \vector{p}$ and numerically compute the sensitivity using finite differences
\begin{equation}
    \frac{d\mathcal{J}}{d\vector{p}}\bigg\rvert_{\vector{p} = \bar{{\vector{p}}}} = \frac{d\mathcal{J}}{dp_{i}}\bigg\rvert_{p_i = \bar{{p}}_i} \approx \frac{\mathcal{J}\big\rvert_{p_i = \bar{p}_i + h} - \mathcal{J}\big\rvert_{p_i = \bar{p}_i}}{h} + O(h), \quad i = 1,2,\dots,n_p,
    \label{eq:finite_difference}
\end{equation}
which needs to be repeated for each $p_i \in \vector{p}$ to determine $d\mathcal{J}/d\vector{p}$. This means that the computational cost of the sensitivity increases linearly with the number of parameters. A more efficient approach is to use the adjoint equation. Analytically, we can express the gradient $d\mathcal{J}/d\vector{p}$ as
\begin{subequations}
    \begin{align}
        \frac{d\mathcal{J}}{d\vector{p}} &= \frac{1}{T}\int_{0}^T \left(\frac{\partial \tilde{\mathcal{J}}}{\partial \vector{p}} + \frac{\partial \tilde{\mathcal{J}}}{\partial \vector{x}}\matrix{Q}\right)dt, \\
        \dot{\matrix{Q}} &= \frac{\partial \vector{F}}{\partial \vector{p}}+\frac{\partial \vector{F}}{\partial \vector{x}}\matrix{Q}, \\
        \matrix{Q}(0) &= \matrix{0},
    \end{align}\label{eq:cont_tangent_linear}
\end{subequations}
where we define $\matrix{Q} \equiv d\vector{x}/d\vector{p},  \;  \matrix{Q}\in \mathbb{R}^{n_x \times n_p}$. We refer to solving the system of equations~\eqref{eq:cont_tangent_linear} as the tangent linear problem. In this case, the sensitivity can be computed by formulating the tangent linear system around a base trajectory and solving for $\matrix{Q}$ in time. Whilst this approach eliminates the finite difference numerical error, it still suffers from a computational cost that is linear with the number of parameters as the dimension of $\matrix{Q}$ grows with the dimension of $\vector{p}$. Adjoint methods circumvent this issue. By introducing the Lagrangian of the objective~\eqref{eq:cont_objective} and the adjoint variables $\vector{q}^+ \in \mathbb{R}^{n_x}$ as the Lagrange multipliers of the constraint~\eqref{eq:cont_constraint}, the gradient $d\mathcal{J}/d\vector{p}$ can be rewritten as
\begin{subequations}
    \begin{align}
        \frac{d\mathcal{J}}{d\vector{p}} &= \frac{1}{T}\int_{0}^T \left(\frac{\partial \tilde{\mathcal{J}}}{\partial \vector{p}} + \vector{q}^{+\top}\frac{\partial \vector{F}}{\partial \vector{p}}\right)dt, \\
        \dot{\vector{q}}^+ &= -\frac{\partial \tilde{\mathcal{J}}}{\partial \vector{x}}^\top -\frac{\partial \vector{F}}{\partial \vector{x}}^\top\revise{\vector{q}^+}, \\
        \vector{q}^+(T) &= 0,
    \end{align}\label{eq:cont_adjoint}
\end{subequations}
where the adjoint variables $\vector{q}^+$ need to solved by integrating the adjoint system backwards in time. The details of the derivation of the adjoint problem from a Lagrangian perspective for a continuous-time system can be found in~\citep{magri2019AdjointMethodsDesign}. In this paper, we derive the tangent linear and adjoint equations for the proposed data-driven algorithm in Appendix~\ref{sec:appendix_adjoint}. 

\subsection{Parameter-aware Echo State Networks}
Given the evolution equations of a dynamical system~\eqref{eq:cont_constraint}, we can compute its adjoint~\eqref{eq:cont_adjoint}. Developing the adjoint solver however requires the system Jacobian, $\partial \vector{F}/\partial \vector{x}$. This Jacobian is code specific. The key idea of this paper is to eliminate code-specific adjoint solvers by having a machine that learns the dynamics from data, and is itself a dynamical system, the Jacobian of which is readily available independent of the system under investigation. Recurrent neural networks (RNNs) are well-suited for this task, as they are neural network architectures that can process sequential data and make forecasts of timeseries from dynamical systems. Although the dynamical systems we consider in Eq.~\eqref{eq:cont_constraint} are continuous in time, RNNs accept data as a sequence in discrete time. RNNs possess memory through a hidden state and thus they can learn temporal relationships. The hidden state enables the RNNs to be framed as discrete dynamical systems~\citep{margazoglou2023StabilityAnalysisChaotic}. In this work, we utilize the Echo State Network (ESN)~\citep{jaeger2004HarnessingNonlinearityPredicting}, which is a type of RNN that is straightforward to train because it does not require backpropagation during training. The ESN is a universal approximator~\citep{grigoryeva2018EchoStateNetworks} and it follows this dynamic equation
 \begin{subequations}
 \begin{equation}\label{eq:esn_step}
     \vector{r}(i+1) = (1-\alpha)\vector{r}(i) + \alpha\tanh(\matrix{W}_{in}\vector{y}_{in}(i)+\matrix{W}\vector{r}(i)),
 \end{equation}
 where $\vector{y}_{in} \in \mathbb{R}^{n_y}$ is the input vector, $\vector{r}(i) \in \mathbb{R}^{n_r}$ is the reservoir state, $\matrix{W}_{in} \in \mathbb{R}^{n_r \times n_y}$ is the input matrix, and $\matrix{W} \in \mathbb{R}^{n_r \times n_r}$ is the state matrix. {Equation~\eqref{eq:esn_step} maps the reservoir state at time $t_i$ to the reservoir state at time $t_{i+1} = t_{i} +\Delta t$.} The prediction of the output is a linear combination of the reservoir state
\begin{equation}\label{eq:esn_readout}
    \vector{\hat{y}}(i+1) = \matrix{W}_{out}\vector{r}(i+1),
\end{equation}   
\end{subequations}
where $\vector{y} \in \mathbb{R}^{n_y}$ is the target output vector, the dynamics of which we aim to model, $\vector{\hat{y}} \in \mathbb{R}^{n_y}$ is the predicted output vector, and $\matrix{W}_{out} \in \mathbb{R}^{n_y \times n_r}$ is the output matrix that maps the reservoir state to the output vector. The matrices $\matrix{W}_{in}$ and $\matrix{W}$ are sparse, randomly generated, and not trained, whilst $\matrix{W}_{out}$ is trained via ridge regression~\cite{lukosevicius2012PracticalGuideApplying}. 
The input matrix is scaled as $\matrix{W}_{in} = \sigma_{in}\tilde{\matrix{W}}_{in}$, where $\tilde{\matrix{W}}_{in}$ is constructed such that each row contains only one nonzero element drawn from a uniform distribution $\sim \mathcal{U}(-1,1)$, i.e., the inputs are activated individually. The state matrix $\matrix{W}$ describes the reservoir connections, and is scaled as $\matrix{W} = \rho \tilde{\matrix{W}}$, where $\tilde{\matrix{W}}$ is an Erd\H{o}s-Renyi matrix with a given average number of nonzero elements per row (connectivity) drawn from a uniform distribution $\sim \mathcal{U}(-1,1)$, \revise{which is  scaled to have a unity spectral radius. So, the scalar $\rho$ is set as the spectral radius of $\matrix{W}$}. In this architecture, the reservoir state variables ``echo'' the input, and so the reservoir holds memory. 
During training $\vector{y}_{in}$ is fed into the network in open-loop and the evolution of the reservoir state is saved. The reservoir state is initialized to a zero vector. An initial time window is used for the washout stage, which eliminates the effects of the initial condition and is discarded in training. The training is performed by solving for $\matrix{W}_{out}$~\citep{lukosevicius2012PracticalGuideApplying}
\begin{equation}
    \matrix{W}_{out}^* = \underset{\matrix{W}_{out}}{\mathrm{arg \, min}}\; \frac{1}{n_y}\sum_{j = 1}^{n_y}\left(\sum_{i = 1}^{N_t}(y_j(i)-\hat{y}_j(i))^2 +\lambda||\vector{w}_{out, j}||^2\right),
\label{eq:wout_optim}
\end{equation}
where $N_t$ is the number of training steps, $\vector{w}_{out, j}$ is the $j^{th}$ row of $\matrix{W}_{out}$, and $\lambda$ is the Tikhonov coefficient. This is a ridge regression problem, which minimizes both the error between network predictions for the next step and the norm of the output weights for regularization. The solution is~\citep{lukosevicius2012PracticalGuideApplying}
\begin{equation}
    \matrix{W}_{out}^* = \matrix{Y}\matrix{R}^\top(\matrix{R}\matrix{R}^\top+\lambda\matrix{I})^{-1},
\label{eq:wout_soln}
\end{equation}
where $\matrix{R} = [\vector{r}(1)\; \vector{r}(2)\; \dots \; \vector{r}(N_t)], \; \matrix{R} \in \mathbb{R}^{n_r \times N_t}$, $\matrix{Y} = [\vector{y}(1) \; \vector{y}(2) \;\dots \; \vector{y}(N_t)], \; \matrix{Y} \in \mathbb{R}^{n_y \times N_t}$, and $\matrix{I}$ is the identity matrix.
Tuning an ESN involves searching for optimal hyperparameters that consist of input matrix scaling $\sigma_{in}$, the spectral radius of the state matrix $\rho$ (typically $\leq 1$ to satisfy the echo state property), the leak rate $\alpha$, and the Tikhonov coefficient $\lambda$. These hyperparameters are determined by validation. A detailed explanation of training and robust validation of Echo State Networks can be found in~\citep{racca2021RobustOptimizationValidation}. After training, the ESN can run autonomously in closed-loop by setting $\vector{y}_{in}(i) = \vector{\hat{y}}(i)$.

In the formulation~\eqref{eq:esn_step}, the ESN is unaware of the system parameters, i.e., it can only operate in one regime. For our purposes, we require an ESN to learn the parameter-dependent dynamics of a system such that we can capture the variations in different regimes. A parameter-aware extension of the ESN has been employed to predict when the system bifurcates to fixed point solutions, a phenomenon known as amplitude death~\cite{xiao2021PredictingAmplitudeDeath}, and bifurcations in multi-stable regimes including chaos~\cite{roy2022ModelfreePredictionMultistability}. It can be formulated as~\citep{xiao2021PredictingAmplitudeDeath}
\begin{equation}
    \vector{r}(i+1) = (1-\alpha)\vector{r}(i) + \alpha\tanh(\matrix{W}_{in}[\vector{y}_{in}(i); \mathrm{diag}(\vector{\bm{\sigma}}_p)(\vector{p}-\vector{k}_p)]+\matrix{W}\vector{r}(i)),
    \label{eq:parameter_esn_step}
\end{equation}
where $(;)$ denotes vertical concatenation, $\vector{p} \in \mathbb{R}^{n_p}$ denotes the system's parameters, which are fed to the network through the input channel and we assume to be constant in time for each system. The input matrix $\matrix{W}_{in} = [\matrix{W}_{in}^y \; \matrix{W}_{in}^p], \; \matrix{W}_{in}^y \in \mathbb{R}^{n_r \times n_y},  \; \matrix{W}_{in}^p \in \mathbb{R}^{n_r \times n_p}$ is constructed such that the parameters are fully connected with all the reservoir nodes, i.e., the weights associated with the parameters $\matrix{W}_{in}^p$ are nonzero for all rows. The parameter-aware ESN adds two additional hyperparameters per physical parameter $p_i$. A scalar $k_{p_i}$ shifts the parameter $p_i$ as $p_i-k_{p_i}$, and a scalar $\sigma_{p_i}$ scales it as $\sigma_{p_i}(p_i-k_{p_i})$. Hence, we apply the input scaling $\sigma_{in}$ on $\matrix{W}_{in}^y$ only. The training of the parameter-aware ESN is performed  as the standard ESN~\eqref{eq:wout_soln}. The dataset contains data from different parameter regimes, which means that we need to save the evolution of the reservoir state for each training regime, and concatenate as $\matrix{R} = [\matrix{R}^1 \; \matrix{R}^2 \; \dots, \matrix{R}^{N_{tr}}]$ (likewise, the outputs are concatenated as $\matrix{Y} = [\matrix{Y}^1 \; \matrix{Y}^2 \; \dots \matrix{Y}^{N_{tr}}]$), where $N_{tr}$ is the number of training regimes.

\section{Adjoint of the Echo State Network}\label{sec:adjoint_esn}
\revise{In this section, we derive the adjoint of the parameter-aware ESN.  In closed-loop, the ESN is governed by, 
\begin{equation}
    \vector{r}(i+1) = (1-\alpha)\vector{r}(i) + \alpha\tanh(\matrix{W}_{in}[\matrix{W}_{out}\vector{r}(i); \mathrm{diag}(\vector{\bm{\sigma}}_p)(\vector{p}-\vector{k}_p)]+\matrix{W}\vector{r}(i)),
    \label{eq:parameter_esn_step_closed_loop}
\end{equation}
which is an autonomous dynamical system. This is the configuration that we use for the computation of the adjoint. We first tackle sensitivity to the system's parameters, and then to the system's initial conditions.} 
\revise{\subsection{Sensitivity to the system's parameters}}
We consider the time-averaged objective functional~\eqref{eq:cont_objective}, which we wish to optimize. In discrete time, we reformulate it as a sum over a discrete number of time steps, denoted by $N$
\begin{equation}
    \mathcal{J} = \frac{1}{N}\sum_{i = 1}^{N} \tilde{\mathcal{J}}(\vector{r}(i)).
\end{equation}
Without loss of generality, we assume that the objective functional does not explicitly depend on the parameters, i.e., $\partial \tilde{\mathcal{J}}/\partial\vector{p} = 0$. For the parameter-aware ESN, the tangent linear problem is
\begin{subequations}
    \begin{align}
        \frac{d\mathcal{J}}{d\vector{p}} &= \frac{1}{N}\sum_{i=1}^{N}\frac{d \tilde{\mathcal{J}}(\vector{r}(i))}{d \vector{r}(i)}\matrix{Q}(i), \\
        \matrix{Q}(i+1) &= \frac{\partial\vector{r}(i+1)}{\partial \vector{p}} + \frac{\partial\vector{r}(i+1)}{\partial \vector{r}(i)}\matrix{Q}(i), \\
        \matrix{Q}(0) &= \matrix{0},
        \label{eq:tangent_initial}
    \end{align}\label{eq:tangent_linear}
\end{subequations}
where $\matrix{Q}(i) \equiv d\vector{r}(i) / d\vector{p}, \; \matrix{Q}(i) \in \mathbb{R}^{n_r \times n_p}$. Physically, Eq.~\eqref{eq:tangent_initial} means that the initial condition is independent of the parameter. Because the dimension of the tangent linear system, $\matrix{Q}(i)$, grows with the number of parameters, adjoint methods enable a computationally cheap and accurate way to obtain sensitivities~\citep{magri2019AdjointMethodsDesign}. Using the Lagrangian formulation of the objective and constraints given by the evolution equations (Appendix~\ref{sec:appendix_adjoint}), the adjoint problem is formulated as
\begin{subequations}
    \begin{align}
        \frac{d\mathcal{J}}{d\vector{p}} &= \sum_{i = 1}^N \vector{q}^{+\top}(i) \frac{\partial \vector{r}(i)}{\partial \vector{p}}, \\
        \vector{q}^+(i) &=  \frac{1}{N}\frac{\revise{d} \tilde{\mathcal{J}}(\vector{r}(i))}{\revise{d} \vector{r}(i)}^\top + \frac{\partial\vector{r}(i+1)}{\partial \vector{r}(i)}^\top\vector{q}^+(i+1),
        \label{eq:adjoint_step} \\
        \vector{q}^+(N) &= \frac{1}{N}\frac{\revise{d} \tilde{\mathcal{J}}(\vector{r}(N))}{\revise{d} \vector{r}(N)}^\top.
        \label{eq:adjoint_terminal}
    \end{align}\label{eq:adjoint}
\end{subequations}
Equation~\eqref{eq:adjoint} is in a form similar to the continuous time equations~\eqref{eq:cont_adjoint}, however, we have replaced the system's Jacobian with the ESN's Jacobian, thus, the adjoint variables are the adjoint of the ESN's reservoir state. Algorithmically, we first obtain the direct solution, $\vector{r}(i)$, as a baseline trajectory by running the ESN in closed-loop for a time window. Then, the adjoint equations~\eqref{eq:adjoint_step} are solved backwards in time starting from the terminal condition $\vector{q}^+(N)$~\eqref{eq:adjoint_terminal}. This procedure requires the computation of the Jacobian, $\partial \vector{r}(i+1)/\vector{r}(i)$, and the partial derivative with respect to the parameters $\partial \vector{r}(i) / \partial \vector{p}$, associated with the ESN. These gradients are provided in Appendix~\ref{sec:appendix_grads}.

The objective functional we are interested in depends on the output vector, e.g.,
\begin{equation}
    \mathcal{J} = \frac{1}{N}\sum_{i = 1}^{N}||\hat{\vector{y}}(i)||^2_2 = \frac{1}{N}\sum_{i = 1}^{N}||\matrix{W}_{out}\vector{r}(i)||^2_2.
    \label{eq:quadratic_objective}
\end{equation}
Thus, the gradient of the objective with respect to the reservoir state is
\begin{equation}
    \frac{\revise{d} \tilde{\mathcal{J}}(\vector{r}(i))}{\revise{d} \vector{r}(i)} = 2\matrix{W}_{out}^\top\matrix{W}_{out}\vector{r}(i).
\end{equation}
Minimization of equation~\eqref{eq:quadratic_objective} defines the optimization problem where the ESN models the physical dynamical system.

\revise{\subsection{Sensitivity to the system's initial conditions}}
\revise{We consider the sensitivity of an objective functional of the final reservoir state after $N$ time steps, i.e., $\mathcal{J} = \mathcal{J}(\vector{r}(N))$ to the initial conditions. In the ESN's equations~\eqref{eq:parameter_esn_step}, the system's initial conditions $\vector{y}(0)$ appear in the input as $\vector{y}_{in}(0)$. When computing the sensitivity to the initial input vector $\vector{y}_{in}(0)$, we need to make a distinction between whether the first step of the propagation of the ESN's equations is performed in open-loop or closed-loop. Because of the recurrence in the ESN, the full state of the ESN is described by the reservoir state. Therefore, for the ESN's estimate of the sensitivity to match the original system, we first run a washout so that the output of the reservoir synchronizes with the original system and then perform the first step in closed-loop. 
(For a use case of the resulting open-loop Jacobian, see an application in data assimilation in~\citep{novoa2024InferringUnknownUnknowns}.) 
The sensitivity  to the initial condition is 
\begin{equation}
        \frac{d\mathcal{J}(\vector{r}(N))}{d\vector{y}_{in}(0)} 
        = \frac{d\mathcal{J}(\vector{r}(N))}{d \vector{r}(N)} \frac{d\vector{r}(N)}{d\vector{r}(1)}\left(\frac{\partial \vector{r}(1)}{\partial \vector{y}_{in}(0)} + \frac{\partial\vector{r}(1)}{\partial\vector{r}(0)}\frac{d\vector{r}(0)}{d\vector{y}_{in}(0)}\right), 
\end{equation}
where the gradients $\partial\vector{r}(1)/\partial\vector{y}_{in}(0)$ and $\partial\vector{r}(1)/\partial\vector{r}(0)$ are found by differentiating the open-loop equation~\eqref{eq:parameter_esn_step} and the gradients $\partial\vector{r}(i+1)/\partial\vector{r}(i), \; i = 1,2,\dots,N-1$ are found by differentiating the closed-loop equation~\eqref{eq:parameter_esn_step_closed_loop}. From the output of the ESN~\eqref{eq:esn_readout}, we find the gradient $d\vector{r}(0)/d\vector{y}_{in}(0) = \matrix{W}_{out}^\dagger$, where $(.)^\dagger$ indicates pseudo-inverse. Alternatively, we compute $d\mathcal{J}(\vector{r}(N))/d\vector{r}(0)$ in closed-loop configuration for all time steps and then determine by the chain rule
\begin{equation}
    \frac{d\mathcal{J}(\vector{r}(N))}{d\vector{y}_{in}(0)} = \frac{d\mathcal{J}(\vector{r}(N))}{d\vector{r}(0)}\matrix{W}_{out}^\dagger. 
\end{equation} 
The tangent linear equations are given by
\begin{subequations}
    \begin{align}
        \frac{d\mathcal{J}(\vector{r}(N))}{d\vector{r}(0)} &=\frac{d\mathcal{J}(\vector{r}(N))}{d \vector{r}(N)}\matrix{Q}(N), \\
        \matrix{Q}(i+1) &= \frac{\partial\vector{r}(i+1)}{\partial \vector{r}(i)}\matrix{Q}(i), \\
        \matrix{Q}(0) &= \matrix{I},
    \end{align}\label{eq:tangent_linear_r0}
\end{subequations}
where $\matrix{Q}(i) \equiv d\vector{r}(i) / d\vector{r}(0), \; \matrix{Q}(i) \in \mathbb{R}^{n_r \times n_r}$ and $\matrix{I}$ is the identity matrix.
The adjoint equations are 
\begin{subequations}
    \begin{align}
        \frac{d\mathcal{J}(\vector{r}(N))}{d\vector{r}(0)} &=  \vector{q}^{+\top}(0), \\
        \vector{q}^+(i) &= \frac{\partial\vector{r}(i+1)}{\partial \vector{r}(i)}^\top\vector{q}^+(i+1),\\
        \vector{q}^+(N) &= \frac{d \mathcal{J}(\vector{r}(N))}{d \vector{r}(N)}^\top.
    \end{align}\label{eq:adjoint_r0}
\end{subequations}
In this case, the number of ``parameters'' is equal to the dimension of the reservoir state vector, which is typically large, and therefore the adjoint approach significantly reduces the computational cost as compared to finite differences or the tangent linear approach.}

\section{Thermoacoustics}\label{sec:thermoacoustics}
We consider a prototypical thermoacoustic system known as the Rijke tube, which consists of a duct with a heat source located inside. The Rijke tube is a setup used to propose fundamental methods of thermoacoustic instabilities because it captures the qualitative nonlinear dynamics and bifurcations observed in real-life applications~\citep[e.g.,][]{juniper2011TriggeringHorizontalRijke,kabiraj2012BifurcationsSelfExcitedDucted,huhn2020StabilitySensitivityOptimisation}. In this model, we make the following assumptions; (i) the acoustics are one-dimensional, (ii) the mean-flow has a low Mach number and the acoustic variables are modelled as perturbations on top of a mean flow with uniform density across the tube, (iii) the heat release is compact, and (iv) the boundary conditions are ideal, i.e., fully reflective. The governing partial differential equations (PDEs) are derived from conservation of momentum, energy, and mass~\citep[e.g.,][]{magri2019AdjointMethodsDesign}
\begin{subequations}
    \begin{align}
        & \frac{\partial u}{\partial t} + \frac{\partial p}{\partial x}= 0, \\
        &\frac{\partial p}{\partial t} + \frac{\partial u}{\partial x} + \zeta p - \dot{q}\delta(x-x_f)= 0,
    \end{align}\label{eq:rijke_pde}
\end{subequations}
where $u$ and $p$ are the acoustic velocity and pressure fluctuations, respectively, $\zeta$ is the frequency dependent damping, which accounts for the acoustic dissipation, $\dot{q}$ is the nonlinear heat release model, and $\delta(x-x_f)$ is a Dirac delta at the heat source location $x_f$. The solution to the governing PDEs is expressed by a Galerkin decomposition on $n_g$ acoustic eigenfunctions as~\citep{zinn1971ApplicationGalerkinMethod}
\begin{equation}
    u(x,t) = \sum_{j = 1}^{n_g}\eta_j(t)\cos(j\pi x), \quad p(x,t) = -\sum_{j = 1}^{n_g}\mu_j(t)\sin(j\pi x), \label{eq:galerkin}
\end{equation}
where $j\pi$ are the non-dimensional angular frequencies of the purely acoustic modes~\citep{magri2014GlobalModesReceptivity}. We denote the vectors of velocity and pressure Galerkin amplitudes by $\vector{\bm{\eta}} = [\eta_1; \; \eta_2; \; \dots; \; \eta_{n_g}]$ and $\vector{\bm{\mu}} = [\mu_1; \; \mu_2; \; \dots; \; \mu_{n_g}]$, respectively. By substituting pressure and velocity variables in the PDEs~\eqref{eq:rijke_pde} with their Galerkin decompositions in~\eqref{eq:galerkin} and projecting the dynamics onto the Galerkin modes, the dynamics of the Galerkin variables $\eta_j$ and $\mu_j$ are described by a $2n_g$ dimensional system of ordinary differential equations (ODEs)
\begin{subequations}
    \begin{align}
    & \dot{\eta}_j-\mu_j j\pi = 0, \label{eq:eta_dot} \\
    & \dot{\mu}_j+\eta_j j\pi+\zeta_j\mu_j+2\dot{q}\sin(j\pi x_f) = 0, \label{eq:mu_dot}
    \end{align}\label{eq:rijke_galerkin_ode}
\end{subequations}
where the modal damping is modelled by $\zeta_j = c_1j^2+c_2j^{1/2}$, and $\dot{q}$ is the heat release rate described by a nonlinear time-delay model~\cite{heckl1990NonlinearAcousticEffects}
\begin{equation}
    \dot{q} = \beta\left(\sqrt{|1+u(x_f,t-\tau)|}-1\right),
    \label{eq:kings}
\end{equation}
where $\beta$ is the heat release strength, and $\tau$ is the time delay between the velocity at the heat source and heat release. The system exhibits different nonlinear solutions, e.g., bifurcation from fixed point to limit cycle, when the parameters $x_f$, $\beta$ and $\tau$ are varied. In the remainder of this paper, we fix $x_f = 0.2$, which causes thermoacoustic instabilities to occur, and analyse the dynamics as functions $\beta$ and $\tau$. The derivative of the nonlinear heat release model~\eqref{eq:kings} with respect to $u(x_f,t-\tau)$ is not defined for $u(x_f,t-\tau) = -1$, which means that the Jacobian and consequently the sensitivity of the system can not be calculated around this value. Therefore, the heat-release model~\eqref{eq:kings} is modified to have smooth dynamics by approximating a small-neighbourhood of $u(x_f,t-\tau) = -1$ with a fourth-degree polynomial \citep{huhn2020StabilitySensitivityOptimisation}.
Further, the time-delayed problem is reformulated as an advection equation via the dummy variable $v$ to make the problem Markovian~\citep{huhn2020StabilitySensitivityOptimisation}
\begin{equation}\label{eq:advection}
    \frac{\partial v}{\partial t} + \frac{1}{\tau}\frac{\partial v}{\partial X} = 0, \quad 0\leq X \leq 1, \quad
    v(X = 0,t) = u(x_f,t).
\end{equation}
The PDE~\eqref{eq:advection} is discretized using a Chebyshev spectral method with $n_c$ points such that the dimension of the system is $2n_g+n_c$ and thus, can be solved with a standard time-marching scheme. For the thermoacoustic system, the objective functional $\mathcal{J}$ is the time-averaged acoustic energy, which relates to the average energy of the undesired oscillations, given by~\citep[e.g.][]{huhn2020StabilitySensitivityOptimisation}
\begin{equation}\label{eq:acoustic_energy}
    E_{ac}(t) = \int_{0}^{1}\frac{1}{2}(u^2(x,t)+p^2(x,t)) dx = \frac{1}{4} \sum_{j = 1}^{n_g}(\eta_j^2(t)+\mu_j^2(t)).
\end{equation}

\section{Thermoacoustic Echo State Network}\label{sec:thermoacoustic_esn}
\begin{figure}
    \includegraphics[width=\linewidth]{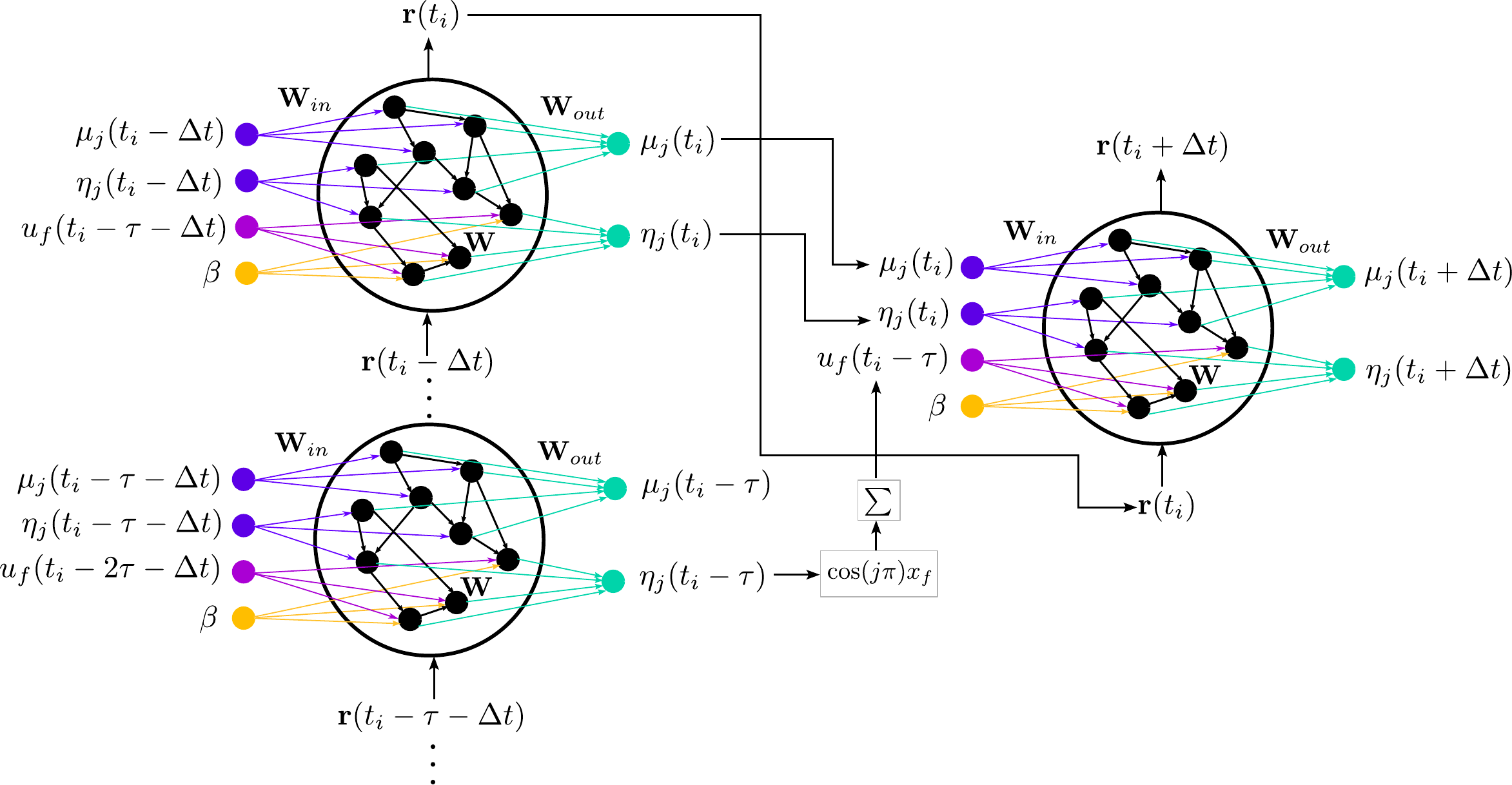}
    \caption{Schematic of the proposed thermoacoustic Echo State Network (T-ESN) in closed-loop configuration. T-ESN hard constrains the physical knowledge about the nonlinear and time-delayed thermoacoustics.}
    \label{fig:thermoacoustic_esn_schematic}
\end{figure}
We propose a parameter-aware ESN with hard-constrained physics for thermoacoustics~\citep{ozan2023HardconstrainedNeuralNetworks}. In the context of machine learning, hard constrainining means embedding prior knowledge into the architecture of the neural network, as opposed to soft-constraining, which refers to imposing prior knowledge as a penalty during training~\citep{doan2020PhysicsinformedEchoState}. This approach makes neural networks more robust and generalizable on unseen data~\citep{ozan2023HardconstrainedNeuralNetworks}. 
First of all, our aim is to develop a parameter-aware ESN that learns the dynamics of the Galerkin amplitudes~\eqref{eq:galerkin} from the thermoacoustic system introduced in Sec.~\ref{sec:thermoacoustics}. Initially, we consider  two cases without any hard constraining: (i) the Galerkin amplitudes as the input and output vectors $\vector{y}_{in}(i) = \hat{\vector{y}}(i) = [\hat{\vector{\bm{\eta}}}(i); \; \hat{\vector{\bm{\mu}}}(i)]$, and (ii) the Galerkin amplitudes and the advection variables as the input and output vectors $\vector{y}_{in}(i) = \hat{\vector{y}}(i) =  [\hat{\vector{\bm{\eta}}}(i); \; \hat{\vector{\bm{\mu}}}(i); \; \hat{\vector{v}}(i)]$. In these cases, we feed both $\beta$ and $\tau$ as parameters through the input channel, $\vector{p} = [\beta; \; \tau]$. However, neither of these options is inherently time delayed. Therefore, we design an ESN for time-delayed systems and refer to it as thermoacoustic ESN (T-ESN). During closed-loop prediction, we compute the delayed velocity at the flame location from past predictions of the velocity Galerkin amplitudes. We then feed the delayed velocity into the network as an input for the prediction of the next time step. One step of T-ESN is
{
\begin{subequations}
\begin{align}
        u_f(i-N_\tau) &= \sum_{j = 1}^{n_g} \hat{\eta}_j(i-N_\tau) \cos(j\pi x_f), \label{eq:delayed_velocity}\\
        \vector{r}(i+1) &= (1-\alpha)\vector{r}(i) + \alpha \tanh(\matrix{W}_{in}[\vector{y}_{in}(i);u_f(i-N_\tau);\sigma_{\beta}(\beta-k_{\beta})]+\matrix{W}\vector{r}(i)), \\
        \hat{\vector{y}}(i+1) &= [\hat{\vector{\bm{\eta}}}(i+1); \; \hat{\vector{\bm{\mu}}}(i+1)] 
        = \matrix{W}_{out}\vector{r}(i+1),
\end{align}\label{eq:thermoacoustic_esn_step}
\end{subequations}
}
where $N_\tau = \mathrm{int}(\tau/\Delta t)$, which implies that the ESN handles delays that are integer multiples of the time step, as it is a discrete-time map. (To extend this, one can use interpolation to approximate the velocity at the desired time delay, however this is beyond the scope of this paper.) A schematic of the T-ESN in closed-loop configuration is shown in Fig.~\ref{fig:thermoacoustic_esn_schematic}. 

The tangent linear problem for computing the gradient with respect to the parameters becomes
\begin{subequations}
    \begin{align}
        \frac{d\mathcal{J}}{d\vector{p}} &=\revise{\frac{1}{N}} \sum_{i = 1}^{N}\frac{\revise{d} \tilde{\mathcal{J}}(\vector{r}(i))}{\revise{d} \vector{r}(i)}\matrix{Q}(i), \\
        \matrix{Q}(i+1) &= \frac{\partial\vector{r}(i+1)}{\partial \vector{p}} + \frac{\partial\vector{r}(i+1)}{\partial \vector{r}(i)}\matrix{Q}(i) + \frac{\partial\vector{r}(i+1)}{\partial \vector{r}(i-N_\tau)}\matrix{Q}(i-N_\tau), \\
        \matrix{Q}(\leq 0) &= \vector{0}.
    \end{align}\label{eq:direct_delayed}
\end{subequations}
The adjoint of the time-delayed system is time-advanced
\begin{subequations}
    \begin{align}
        \frac{d\mathcal{J}}{d\vector{p}} &= \sum_{i = 1}^N \vector{q}^{+\top}(i) \frac{\partial \vector{r}(i)}{\partial \vector{p}}, \\
        \vector{q}^+(i) &= \frac{1}{N}\frac{\revise{d} \tilde{\mathcal{J}}(\vector{r}(i))}{\revise{d}\vector{r}(i)}^\top + \frac{\partial\vector{r}(i+1)}{\partial \vector{r}(i)}^\top\vector{q}^+(i+1) + \frac{\partial\vector{r}(i+1+N_\tau)}{\partial \vector{r}(i)}^\top\vector{q}^+(i+1+N_\tau), \\
        \vector{q}^+(N) &= \frac{1}{N}\frac{\revise{d} \tilde{\mathcal{J}}(\vector{r}(N))}{\revise{d}\vector{r}(N)}^\top, \\
        \vector{q}^+(\geq N) &= \vector{0}.
    \end{align}\label{eq:adjoint_delayed}
\end{subequations}
The derivation of the time-delayed adjoint is provided in Appendix~\ref{sec:appendix_adjoint}. The time-delayed Jacobian is provided in Appendix~\ref{sec:appendix_grads}. \revise{(The tangent linear and adjoint equations for the initial condition sensitivity are found similarly by including the time-delayed or time-advanced Jacobian to the Eqs.~\eqref{eq:tangent_linear_r0}~and~\eqref{eq:adjoint_r0} in Sec.~\ref{sec:adjoint_esn}, respectively.)}

To obtain the sensitivity to time delay, we need 
\begin{equation}
    \frac{\partial \vector{r}(i+1)}{\partial \tau} =  \frac{\partial \vector{r}(i+1)}{\partial u_f(i-\tau)} \frac{\partial u_f(i-\tau)}{\partial \tau}
\end{equation}
In the delayed formulation of the T-ESN, the delayed velocity $u_f(t-\tau)$~\eqref{eq:delayed_velocity} does not continuously depend on $\tau$. We approximate the gradient $\partial u_f(t-\tau)/\partial \tau$ with a central finite difference
\begin{equation}
    \frac{\partial u_f(i-\tau)}{\partial \tau} \approx \frac{u_f(i - (N_\tau+1))- u_f(i -(N_\tau-1))}{2\Delta t}.
\end{equation}

Motivated by thermoacoustic physics, we design the input weights matrix based on the following observations; (i) the acoustics are linear and the modes are sparsely connected, (ii) the heat release is nonlinear, (iii) the heat release strength $\beta$ is only coupled with $u_f(t-\tau)$
\begin{equation}
    \matrix{W}_{in} = \begin{bmatrix}
        \vector{w}_{1,1} & 0 & 0 & 0 & 0 & 0\\
        0 & \vector{w}_{2,2} & 0 & 0 & 0 & 0\\
        0 & 0 & \ddots & 0 & 0 & 0\\
        0 & 0 & 0 & \vector{w}_{2n_g,2n_g} & 0 & 0 \\
        0 & 0 & 0 & 0 & \vector{w}_{f} & \vector{w}_{\beta} 
    \end{bmatrix},
\end{equation}
where $\vector{w}_{j,j} \in \mathbb{R}^{\mathrm{int}(n_r/(2n_g+1))}, \; j = 1,\dots,2n_g, \vector{w}_{f} \in \mathbb{R}^{\mathrm{int}(n_r/(2n_g+1))}, \vector{w}_{\beta} \in \mathbb{R}^{\mathrm{int}(n_r/(2n_g+1))}$ The $\tanh$ activation acts approximately linearly near zero, so the terms that linearly contribute to the dynamics have a smaller scaling, whereas the nonlinear terms have a larger scaling. Therefore, we scale the input weights that multiply the linear acoustics with one scaling i.e., $\vector{w}_{j,j} = \sigma_{in}\tilde{\vector{w}}_{j,j}$, and the input weights that multiply the delayed velocity that contributes to the nonlinear heat release term with another scaling, i.e., $\vector{w}_{f} = \sigma_{f}\tilde{\vector{w}}_{f}$, because we expect that $\sigma_{f} > \sigma_{in}$ ($\tilde{\vector{w}}_{j,j}$, $\tilde{\vector{w}}_{f}$ and $\vector{w}_{\beta}$ are drawn from the uniform distribution $\sim \mathcal{U}(-1,1)$ as before). While doing so we do not impose the linearity, but rather allow the machine to discover this via the optimization of the scaling terms. Separately, the parameter $\beta$ is shifted and scaled by $\sigma_{\beta}(\beta-k_{\beta})$. The scalings are determined by the hyperparameter search. In summary, we hard constrain the physical knowledge about the thermoacoustics in the network architecture in two ways: (i) by making the network explicitly time-delayed, and (ii) by designing the input weights matrix.

\section{Learning thermoacoustics and parameter sensitivities}\label{sec:results_learn}
{In this section, we learn the dynamics of the Rijke tube parameterized by the heat release strength, $\beta$, and the time delay, $\tau$. First, we focus on limit cycle regimes and find that the thermoacoustic ESN (T-ESN) outperforms other parameter-aware ESN architectures in terms of generalizability over unseen regimes. Then, using the trained T-ESN, we infer the adjoint sensitivities of the acoustic energy to these parameters. We compare the time series predictions and inferred adjoint sensitivities to the ground truth. The ground truth is obtained by integrating the original system described by the ODEs~\eqref{eq:rijke_galerkin_ode}, and its adjoint equations~\eqref{eq:cont_adjoint}, which requires the computation of the Jacobian of~\eqref{eq:rijke_galerkin_ode}.}
\subsection{Training and validation}\label{sec:train_and_val}
For training and validation, we create a dataset from time-delays, $\tau = \{0.1, 0.15, 0.2, 0.25, 0.3\}$ and heat release strengths, $\beta = \{1.0, 2.0, 3.0, 4.0, 5.0\}$, giving a total of 25 thermoacoustic regimes. These parameters generate limit cycle solutions except for $(\beta = 5.0, \tau = 0.15)$, which produces a quasiperiodic solution. We fix the flame location $x_f = 0.2$, modal damping coefficients $c_1 = 0.1$ and $c_2 = 0.06$, the number of Galerkin modes $n_g = 4$, and the number of Chebyshev discretization $n_c = 10$~\citep{huhn2020StabilitySensitivityOptimisation}. With four Galerkin modes, we observe the rich nonlinear behaviour associated with thermoacoustics. We generate the data by time-marching the ODEs~\eqref{eq:rijke_galerkin_ode} using the $\texttt{scipy.odeint}$ tool with a time step of 0.001, which we \revise{downsample} to 0.01 for the ESN. We remove a transient of 200 time units, select the next 4 time units as washout and 8 time units as training data. (We found that increasing the training length further does not improve the performance whereas it increases the computational effort.) We randomly split this dataset into 20 regimes for training and 5 regimes for validation of the hyperparameters. During validation, we evaluate the closed-loop performance of 5 random realisations of the ESN, i.e., realisations of the randomly generated input and state matrices, over 5 trajectories, or folds, of length 4 time units (approximately 2 periods) starting from random time steps. As error metric, we compute relative $\ell_2$-error between the prediction, $\hat{\vector{y}}$ and the ground truth $\vector{y}$ 
\begin{equation}\label{eq:rel_l2}
    \epsilon = \frac{\sqrt{\sum_{i=1}^{N_v}\sum_{j=1}^{n_y}(y_j(i)-\hat{y}_j(i))^2}}{\sqrt{\sum_{i=1}^{N_v}\sum_{j=1}^{n_y}y_j^2(i)}},
\end{equation}
where $N_v$ is the number of validation steps, which we average over the number of folds, number of validation regimes, and number of realisations. We employ a Bayesian optimization scheme using the $\texttt{skopt.gp\_minimize}$ tool to determine the set of hyperparameters that have the minimal validation error~\cite{racca2021RobustOptimizationValidation}. 

We consider three networks: (i) the standard parameter-aware ESN~\eqref{eq:parameter_esn_step} with the Galerkin amplitudes as the input and output vectors $\vector{y}_{in}(i) = \hat{\vector{y}}(i) = [\hat{\vector{\bm{\eta}}}(i); \; \hat{\vector{\bm{\mu}}}(i)]$, (ii) we augment the input and output vectors with the advection variable $\vector{y}_{in}(i) = \hat{\vector{y}}(i) =  [\hat{\vector{\bm{\eta}}}(i); \; \hat{\vector{\bm{\mu}}}(i); \; \hat{\vector{v}}(i)]$, and (iii) thermoacoustic ESN (T-ESN)~\eqref{eq:thermoacoustic_esn_step}. The standard and the advection-augmented parameter-aware ESNs have both $\beta$ and $\tau$ as explicit parameters in the input channel, i.e., in Eq.~\eqref{eq:parameter_esn_step} $\vector{p} = [\beta; \; \tau]$. The T-ESN has $\beta$ as a parameter in the input channel and $\tau$ is modelled as a time-delay in the delayed velocity term in the input channel $u_f(t-\tau)$~\eqref{eq:delayed_velocity}. For comparison, we fix the reservoir size, \revise{i.e., number of reservoir state variables $n_r$,} to 1200 and connectivity of the reservoir state variables, \revise{i.e., average number of nonzero elements per row in the state matrix $\matrix{W}$}, to 20. The optimal hyperparameters are provided in Table~\ref{tab:hyperparameters}. The first row of Table~\ref{tab:hyperparameters} shows the ranges from which the hyperparameters are sampled during the validation; $\rho$, $\sigma_{in}$, $\sigma_{f}$, $\sigma_{\beta}$, $\sigma_{\tau}$, $\alpha$ and $\lambda$ are sampled uniformly on a $\log_{10}$ scale, while $k_{\beta}$ and $k_{\tau}$ are sampled uniformly.

\begin{table}
    \caption{Optimal hyperparameters of (i) standard parameter-aware ESN~\eqref{eq:parameter_esn_step}, (ii) advection-augmented parameter-aware ESN~\eqref{eq:parameter_esn_step}, (iii) thermoacoustic ESN~\eqref{eq:thermoacoustic_esn_step} trained on $\beta = \{1.0, 2.0, 3.0, 4.0, 5.0\}$ and $\tau = \{0.1, 0.15, 0.2, 0.25, 0.3\}$ regimes. Refer to the Eqs.~\eqref{eq:parameter_esn_step}, and~\eqref{eq:thermoacoustic_esn_step} for the hyperparameters.}
    
    \begin{tabular}{c|c|c|c|c|c|c|c|c|c}
                    &  $\rho$ & $\sigma_{in}$ & $\sigma_{f}$ & $\sigma_{\beta}$ & $k_{\beta}$ &$\sigma_{\tau}$ & $k_{\tau}$ & $\alpha$ & $\lambda$ \\ \hline
                    &  0.01 -- 1.0 & 0.01 -- 2.0 &  0.01 -- 2.0 & 0.01 -- 2.0 & -10.0 -- 10.0 & 0.01 -- 2.0 & -1.0 -- 1.0 & 0.01 -- 1.0 &  $10^{-6}$ -- $10^{-1}$ \\
        ESN (i)     & 0.0363 & 0.3476 & - & 0.0152 & -5.3759 & 0.7741 & -0.2063 & 0.0597 & $10^{-5}$ \\
        ESN (ii)    & 0.0367 & 0.3293 & - & 0.0162 & -5.4293 & 0.7319 & 0.72734 & 0.0627 & $10^{-5}$ \\
        ESN (iii)   & 0.0125 & 0.0790 & 1.0501 & 0.1092 & -10.0 & - & - & 0.0363 & $10^{-6}$
    \end{tabular}\label{tab:hyperparameters}
\end{table}

\subsection{Prediction performance}\label{sec:pred_perf}
\begin{figure}
    \includegraphics[width=\linewidth]{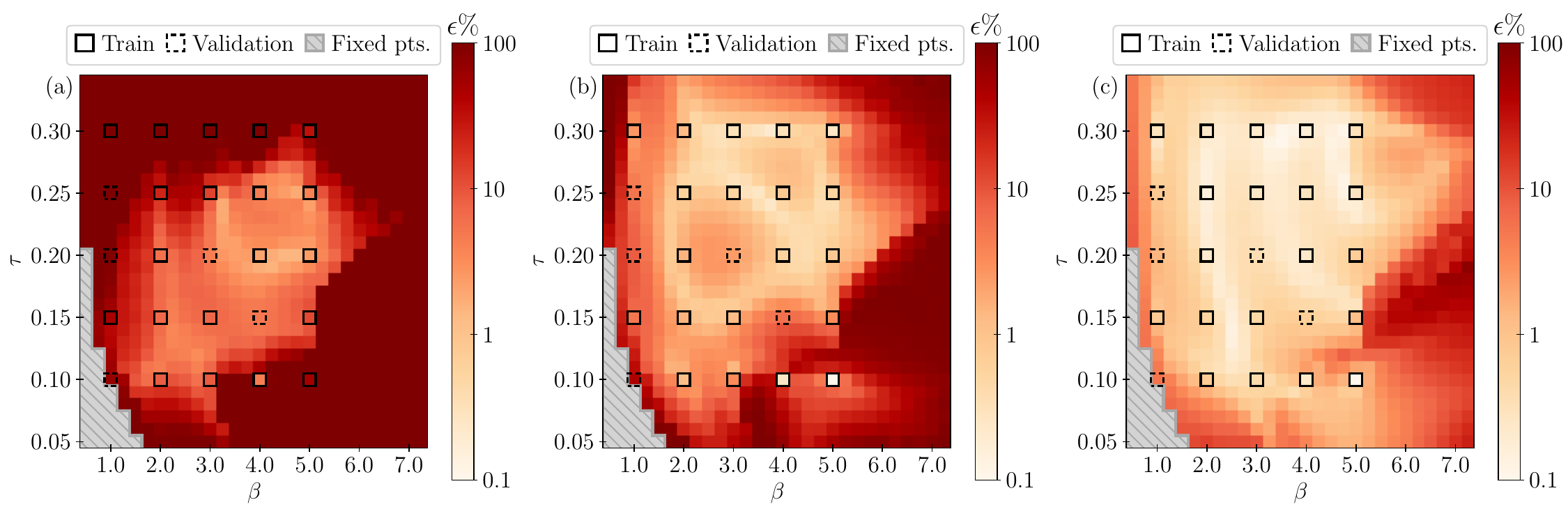}
    \caption{Short-term \revise{closed-loop} prediction performance of (a) standard parameter-aware ESN~\eqref{eq:parameter_esn_step}, (b) advection-augmented parameter-aware ESN~\eqref{eq:parameter_esn_step}, (c) thermoacoustic ESN~\eqref{eq:thermoacoustic_esn_step}. Relative $\ell_2$ error between prediction and ground truth of $[\vector{\bm{\eta}}; \; \vector{\bm{\mu}}]$ over 20 time units on a $\log_{10}$ scale.}\label{fig:ensemble_predict}
\end{figure}

The computed adjoint sensitivities are only as accurate as the model we have. Therefore, the first task of the ESN is to accurately emulate the dynamics of regimes both seen and unseen during training. This includes two subtasks: (i) time-accurate prediction from a given initial condition, and (ii) convergence to the correct attractor. The latter is necessary for long-term performance because we are interested in the sensitivity of long-time averaged quantities. Figure~\ref{fig:ensemble_predict} shows the short-term prediction performance of the standard, advection-augmented and thermoacoustic ESNs over a grid of $\beta$ and $\tau$ values. For a given regime, after a washout stage, the ESN is run in closed-loop for prediction. This means that the network is forecasting the dynamics. 
The relative $\ell_2$ error~\eqref{eq:rel_l2} between prediction and ground truth of $[\vector{\bm{\eta}}; \; \vector{\bm{\mu}}]$ over 20 time units is averaged over 5 folds starting from random time steps, and 5 realisations of the ESN. Standard parameter-aware ESN is not capable of sustaining the limit cycle solutions for approximately 10 periods even for the training regimes, whereas the augmentation of the advection variables $\vector{v}$ in the input and output significantly improves the performance. \revise{The T-ESN} generalizes to unseen regimes. Owing to the nonlinearity in the system, the prediction performance decreases as we move away from the training regimes.

\begin{figure}
    \includegraphics[width=\linewidth]{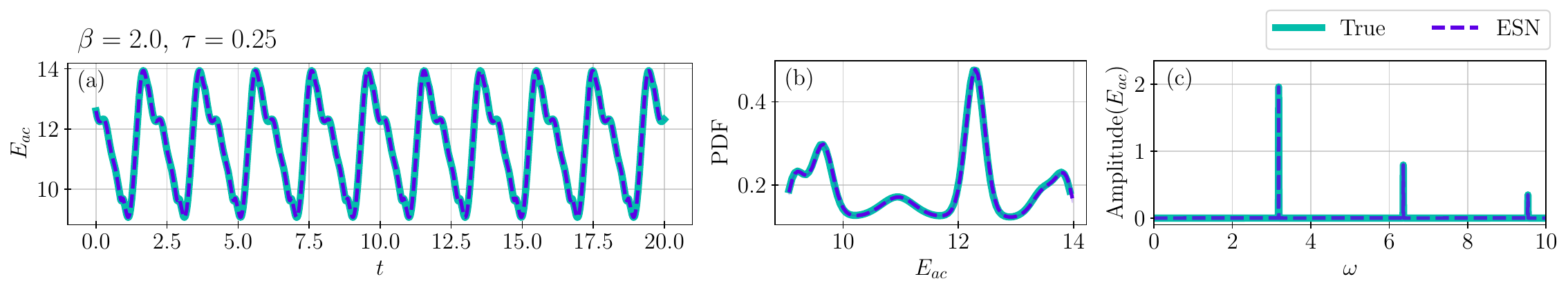}
    \includegraphics[width=\linewidth]{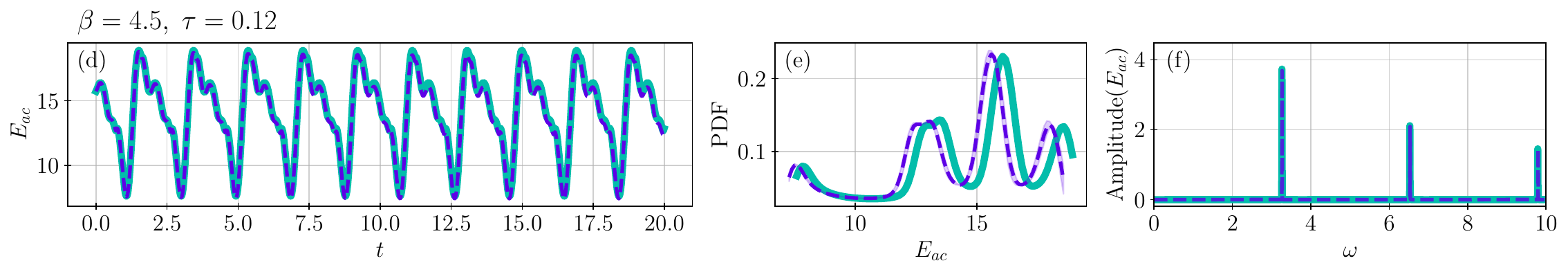}
    \caption{\revise{Thermoacoustic Echo State Network learns limit cycle regimes both seen and unseen during training. Short-term prediction, probability distribution function (PDF), and amplitude spectrum of the acoustic energy of (a-c) a limit cycle regime seen during training ($\beta = 2.0, \; \tau = 0.25$) and (d-f) an unseen limit cycle regime ($\beta = 4.5, \; \tau = 0.12$). In the PDFs, the mean of an ensemble of 5 realisations of the ESN is shown, one standard deviation from the mean is plotted in lighter colour.}}\label{fig:lco}
\end{figure}

In Fig.~\ref{fig:lco}, we show the performance of the T-ESN on (i) a limit cycle regime seen during training ($\beta = 2.0, \; \tau = 0.25$), and (ii) an unseen limit cycle regime ($\beta = 4.5, \;\tau = 0.12$). Based on Fig.~\ref{fig:ensemble_predict}, we choose regime (ii) as an example of an unseen regime with higher relative $\ell_2$ error compared to the other limit cycle regimes. In these figures, we assess the performance by (from left to right) short-term time-series prediction for 20 time units after washout, the probability density function (PDF), i.e., statistics, and amplitude spectrum of the acoustic energy $E_{ac}$~\eqref{eq:acoustic_energy}. 
For the PDF and amplitude spectrum, we feed the initial condition of zeros except $\eta_1(t_0) = 1.0$ in the washout stage, let the ESN evolve in closed-loop, discard a transient of 200 time units and use the next 1000 time units. This procedure and results verify that the ESN has learned the correct limit cycle attractor because it converges to this set of states from an initial condition that is not in the training set. 

\subsection{Computation of adjoint sensitivities}\label{sec:results_adj}
\begin{figure}
    \includegraphics[width=\linewidth]{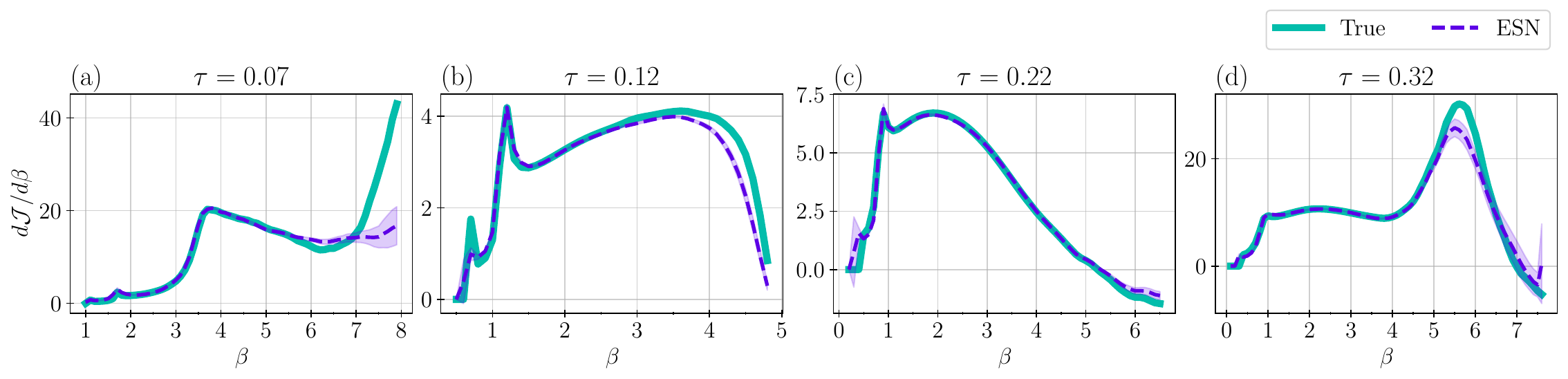}
    \caption{\revise{Thermoacoustic Echo State Network infers sensitivity of the time-averaged acoustic energy, i.e., objective $\mathcal{J}$, to heat release strength $\beta$. Shown $d\mathcal{J}/d\beta$ at fixed (a) $\tau = 0.07$, (b) $\tau = 0.12$, (c) $\tau = 0.22$, (d) $\tau = 0.32$. The mean of the sensitivity of an ensemble of 5 realisations of the ESN is shown, one standard deviation from the mean is plotted in lighter colour.}}\label{fig:sensitivity_beta}
\end{figure}
\begin{figure}
    \includegraphics[width=\linewidth]{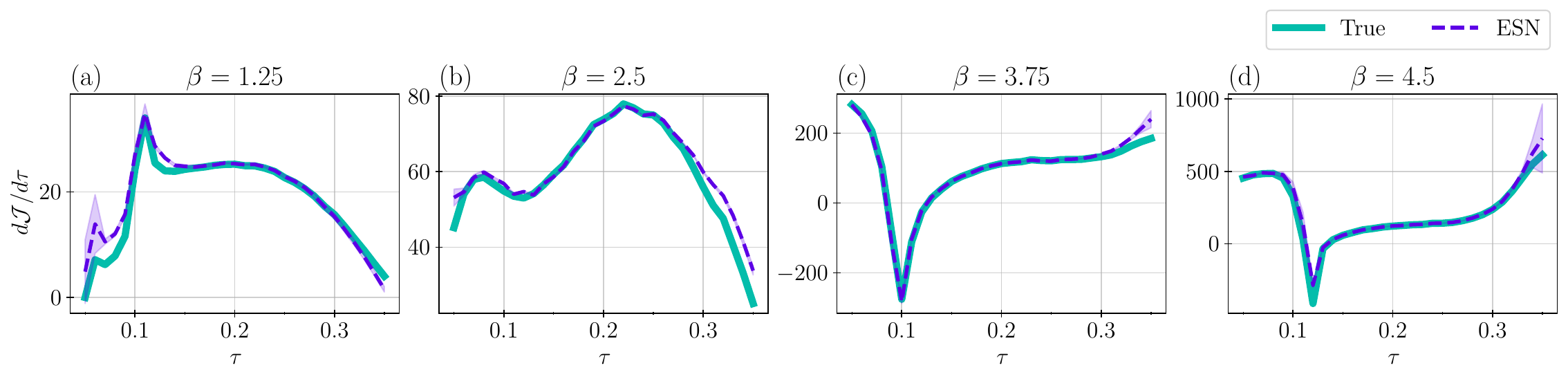}
    \caption{\revise{Thermoacoustic Echo State Network infers sensitivity of the time-averaged acoustic energy, i.e., objective $\mathcal{J}$, to time-delay $\tau$. Shown $d\mathcal{J}/d\tau$ at fixed (a) $\beta = 1.25$, (b) $\beta = 2.5$, (c) $\beta = 3.75$, (d) $\beta = 4.5$. The mean of the sensitivity of an ensemble of 5 realisations of the ESN is shown, one standard deviation from the mean is plotted in lighter colour.}}\label{fig:sensitivity_tau}
\end{figure}
We compare the adjoint sensitivities computed by the T-ESN following the procedure described in Sec.~\ref{sec:adjoint_esn} with the adjoint sensitivities obtained from the original system described by Eq.~\eqref{eq:rijke_galerkin_ode}. We refer to the latter as the ground truth, which is computed by deriving the Jacobian of the ODEs~\eqref{eq:rijke_galerkin_ode} and integrating the adjoint equations~\eqref{eq:cont_adjoint}. The sensitivities are shown for varying $\beta$ and $\tau$ values in \revise{Figs.~\ref{fig:sensitivity_beta}~and~\ref{fig:sensitivity_tau}, respectively}. For ease of results, we show the sensitivity to $\beta$ when $\beta$ is varied and $\tau$ is fixed, and vice versa. In the washout stage, we feed the initial condition of zeros except $\eta_1(t_0) = 1.5$, assuming that no data is available from the original system for the washout. This value of $\eta_1(t_0)$ makes sure that the trajectories converge to the stable limit cycle and not the stable fixed point in the multistable regimes. After the washout and the transient, we solve the adjoint equations around a trajectory of 100 time units, for which we find that the adjoint sensitivity converges to a steady-state value. The learned adjoint sensitivities match closely with the ground truth, while the results are also robust to different realisations of the network. Even though the training data was generated from discrete values of $(\beta,\tau)$ on a grid (Fig.~\ref{fig:ensemble_predict}), we are able to obtain sensitivity information from a range of $(\beta,\tau)$ values that are unseen during the training using a single instance of the T-ESN. The smoothness of the sensitivity curve also indicates that the dynamics of the T-ESN varies smoothly with the parameters. The accuracy decreases as we move away from the training regimes, e.g., beyond $\beta = 5.0$ in \revise{the $d\mathcal{J}/d\beta$} plot at $\tau = 0.32$ \revise{(Fig.~\ref{fig:sensitivity_beta}~(d))}, but the ESN captures the trend of the sensitivities. 

\revise{Figs.~\ref{fig:sensitivity_beta}~and~\ref{fig:sensitivity_tau} also show} that the T-ESN infers the existence of bifurcations to fixed point solutions. The fixed point of the Rijke tube is the zero vector, i.e., the mean flow, which means that the sensitivity is zero for the fixed point regimes, e.g., at fixed $\tau = 0.22$ (Fig.~\ref{fig:sensitivity_beta}~(c)) for $\beta < 0.45$. Albeit the parameter value at which the bifurcation occurs might have a small error compared to the true system. For example, at fixed $\tau = 0.22$ the Hopf bifurcation occurs at $\beta = 0.45$, whereas 4 out of 5 realisations of the T-ESN predict it as $\beta \approx 0.35$, and the remaining realisation predicts it as $ \beta \approx 0.30$. 

\section{Gradient-based optimization}\label{sec:results_grad}
\begin{figure}
    \includegraphics[width=\linewidth]{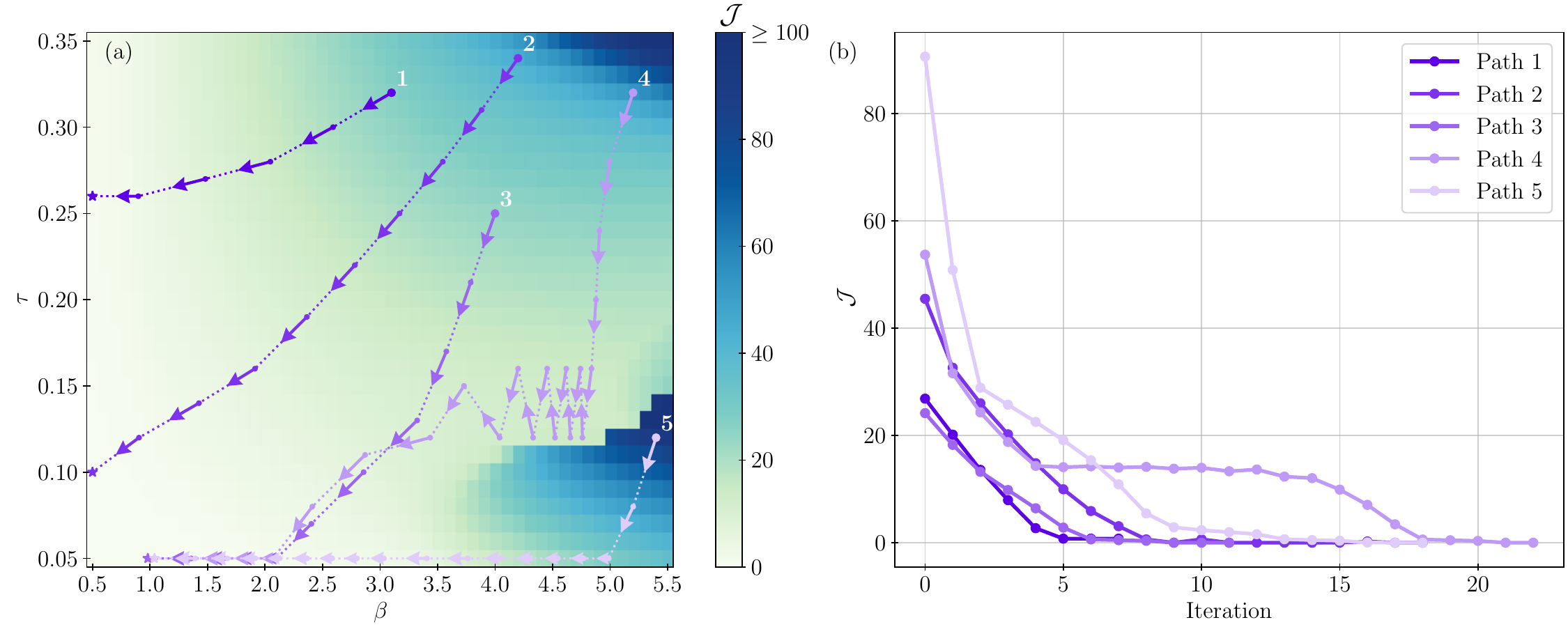}
    \caption{Gradient-based optimization of the time-averaged acoustic energy starting from different initial $(\beta, \tau)$ using adjoint sensitivities computed by a single instance of the thermoacoustic Echo State Network. In (a), example optimization paths starting from different initial $(\beta,\tau)$ are plotted on top of the landscape of the objective functional $\mathcal{J}$. The values $\mathcal{J}$ takes during the iterations of optimization are plotted in (b).}\label{fig:energy_grid}
\end{figure}

\begin{figure}
    \includegraphics[width=\linewidth]{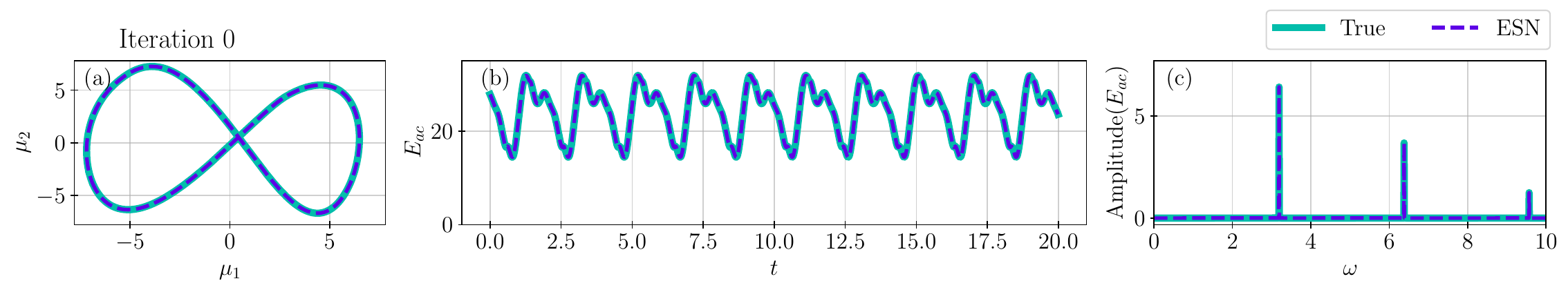}
    \includegraphics[width=\linewidth]{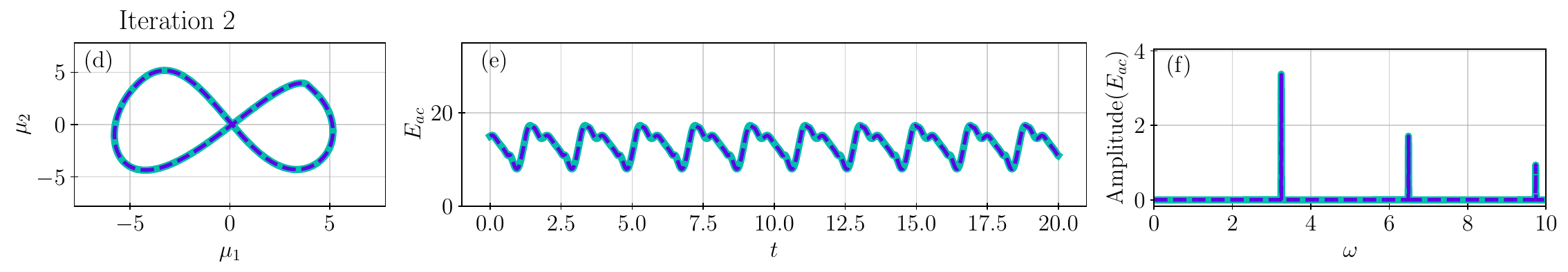}
    \includegraphics[width=\linewidth]{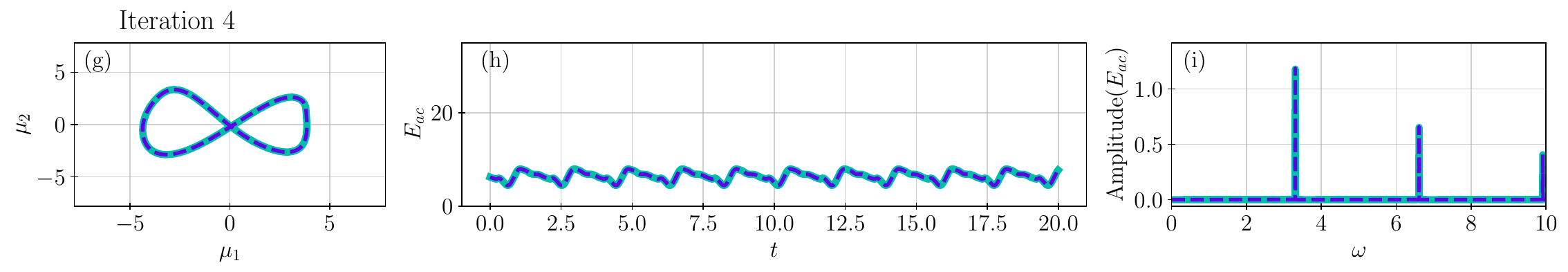}
    \includegraphics[width=\linewidth]{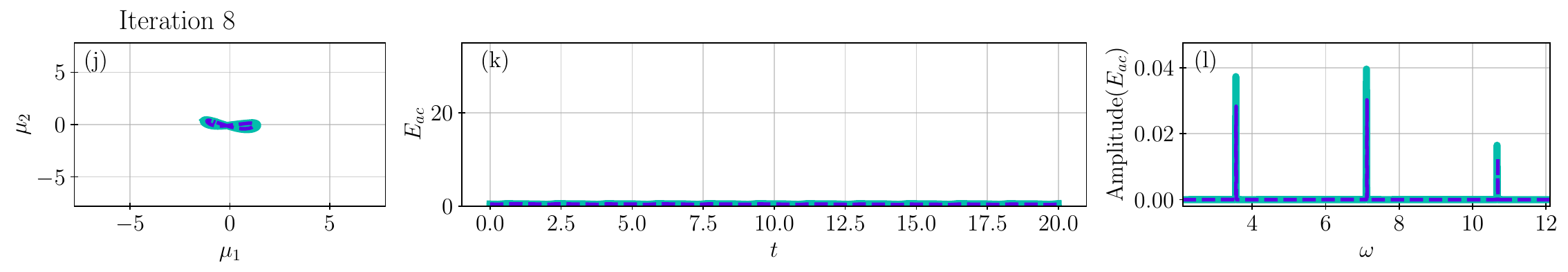}
    \caption{Gradient-based optimization of time-averaged acoustic energy marked as Path 3 in Fig.~\ref{fig:energy_grid}. Shown (a-c) iteration 0; $(\beta = 4.0, \tau = 0.25)$, (d-f) iteration 2; $(\beta = 3.57, \tau = 0.17)$, (g-i) iteration 4; $(\beta = 2.85, \tau = 0.1)$, and (j-l) iteration 8; $(\beta = 1.64, \tau = 0.07)$. From left to right; the first two pressure modes in the phase space, short-term timeseries prediction of the acoustic energy, and long-term amplitude spectrum of the acoustic energy. The optimization follows a path along which the amplitude of the oscillations and the acoustic energy decreases, while the thermoacoustic Echo State Network accurately predicts the changing dynamics.}\label{fig:optimization}
\end{figure}

We apply the inferred adjoint sensitivities to a gradient-based optimization framework. {One realisation of the T-ESN suffices for the computation of the gradients and optimization for all parameters.} In this work, the goal of the optimization is to determine the set of parameters $(\beta,\tau)$ that minimizes the time-averaged acoustic energy~\eqref{eq:acoustic_energy}. We restrict the domain to $\beta = [0.5, 5.5], \; \tau = [0.05, 0.35]$, which prevents the parameters to take negative (non-physical) values, and the system to bifurcate to quasiperiodic or chaotic regimes. We wish to avoid this during gradient-based optimization because first, the acoustic energy tends to display discontinuous jumps at the parameter values where the system bifurcates \citep{huhn2020StabilitySensitivityOptimisation}, which can lead to undefined gradients between different nonlinear regimes. Second, while the adjoint sensitivity can be computed for limit-cycle and quasiperiodic regimes for arbitrary integration times, for chaotic regimes the integration time is limited before the adjoint system becomes unstable, and the sensitivities diverge~\citep{lea2000SensitivityAnalysisClimate,huhn2020StabilitySensitivityOptimisation}. \revise{We deal with the sensitivity of the chaotic regimes separately in Sec.~\ref{sec:results_chaos}.} Another consideration is the respective orders of magnitudes of the parameters and the sensitivities to them. From Figs.~\ref{fig:sensitivity_beta}~and~\ref{fig:sensitivity_tau}, $\beta$ and $\tau$ as well as $d\mathcal{J}/d\beta$ and $d\mathcal{J}/d\beta$ have different orders of magnitudes, i.e., $\beta\sim O(1), \; \tau\sim O(10^{-1})$. The difference in the orders of magnitudes requires the parameters and their respective gradients to be normalized, such that $\breve{\beta} = \beta / \bar{\beta}, \; \breve{\tau} = \tau / \bar{\tau}, \; d\mathcal{J}/d\breve{\beta} = \bar{\beta}d\mathcal{J}/d\beta, \; d\mathcal{J}/d\breve{\tau} = \bar{\tau}d\mathcal{J}/d\tau$, where $\bar{(\cdot)}$ denotes the mean of the parameter in the domain considered~\citep{huhn2020StabilitySensitivityOptimisation}. The optimization is performed with the steepest-descent method using the normalized gradients with a step size of 0.2. (Since $\tau$ can only take discrete values, we round it according to the time-step size of the ESN before the next step of the optimization.) The optimization stops when (i) the acoustic energy is near zero, $\mathcal{J}_{i} < 10^{-4}$, or (ii) the relative change in acoustic energy is below a threshold, $(\mathcal{J}_{i}-\mathcal{J}_{i-1})/(\mathcal{J}_{i-1}) < 10^{-4}$ where subscript $i$ denotes the $i^{th}$ iteration step. Iterations of optimization starting from different initial $(\beta, \tau)$ are shown in Fig.~\ref{fig:energy_grid}. The optimal $(\beta, \tau)$ values are found to be in the fixed point regimes, i.e., when the instability is suppressed. For the Rijke tube, the fixed point is the zero vector (mean flow), and as such the acoustic energy, i.e., the acoustic energy of the fluctuations from the mean flow, is equal to zero at these values. This result shows the significance of the ability of the ESN to infer fixed points. The process of optimization is illustrated for Path 3 in Fig~\ref{fig:optimization}.  The optimization follows a path where the amplitude of the oscillations and so the acoustic energy decreases, for which the ESN accurately predicts the changing dynamics. Similar behaviour is observed for other paths. An exceptional case is Path 4, which passes through a saddle in the landscape of the objective function near $(\beta = 4.5, \tau = 0.15)$. The change in the sign of the gradient, as shown in Fig.~\ref{fig:sensitivity_tau}~(d) in the sensitivity plot for $\beta = 4.5$ around $\tau = 0.13$, causes the ``zigzagging'' path in this region. 

\section{Robustness to noise}\label{sec:results_noise}
\begin{figure}
    \includegraphics[width=0.5\linewidth]{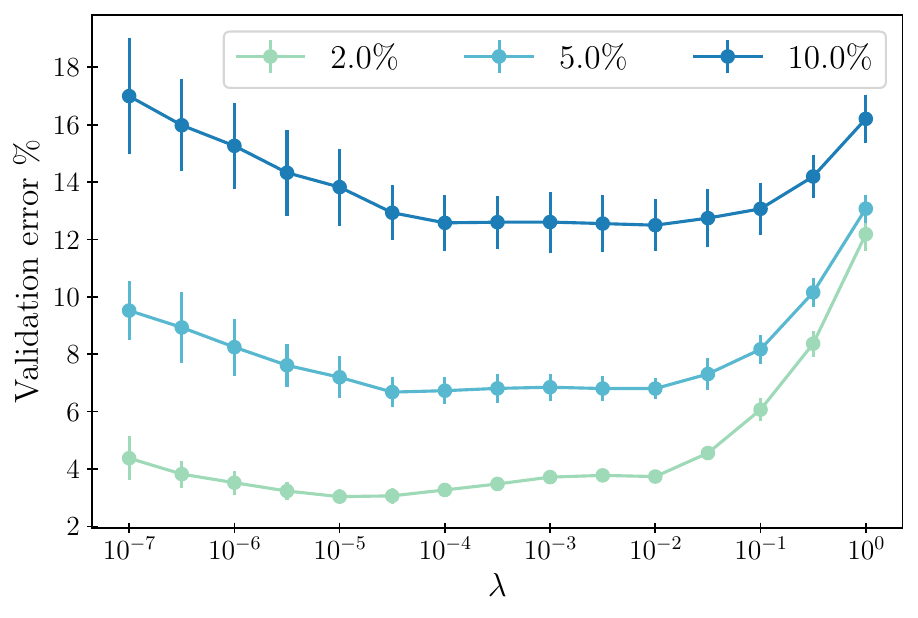}
    \caption{\revise{Selection of the regularization parameter when the data is noisy. Shown is the validation error of thermoacoustic Echo State Network on data with different levels of noise (indicated in the legend) when the training is performed with increasing regularization. The error statistics for 20 different realisations of noise are plotted.}}\label{fig:noise_validation}
\end{figure}

If the data is acquired from real-life experiments, the measurements may be corrupted by noise. This is the case when the system equations are unknown and we wish to learn sensitivities from experimental data. To test the performance of the T-ESN under such a scenario, we consider zero-mean additive Gaussian noise with a standard deviation that is equal to \revise{2\%, 5\%, and 10\%} of the standard deviation of the true data (\revise{these noise levels correspond to 33, 26, and 20 dB signal-to-noise ratios, respectively}). We repeat the training on the noisy data using the hyperparameters chosen before (Tab.~\ref{tab:hyperparameters}) and \revise{investigate the effect of  regularization. The regularization parameter, $\lambda$ can be determined by evaluating the validation error during hyperparameter selection. Fig.~\ref{fig:noise_validation} shows the statistics (mean and one standard deviation) of the validation error of the T-ESN for different levels of noise. The validation error is given by the mean relative $\ell_2$ error~\eqref{eq:rel_l2} as described in Sec.~\ref{sec:train_and_val}. (In the noisy case, we increase the number of evaluated trajectories to 20 as this helps with the covergence of the error statistics.) To obtain the error statistics, we repeat this calculation for 20 different realisations of noise applied to the training and validation data. Because the validation data  contains noise, the irreducible error is the noise level. We find that there is a range of $\lambda$ values that results in a similarly low validation error, and this optimal range of regularization increases with increasing levels of noise.}

\begin{figure}
    \includegraphics[width=\linewidth]{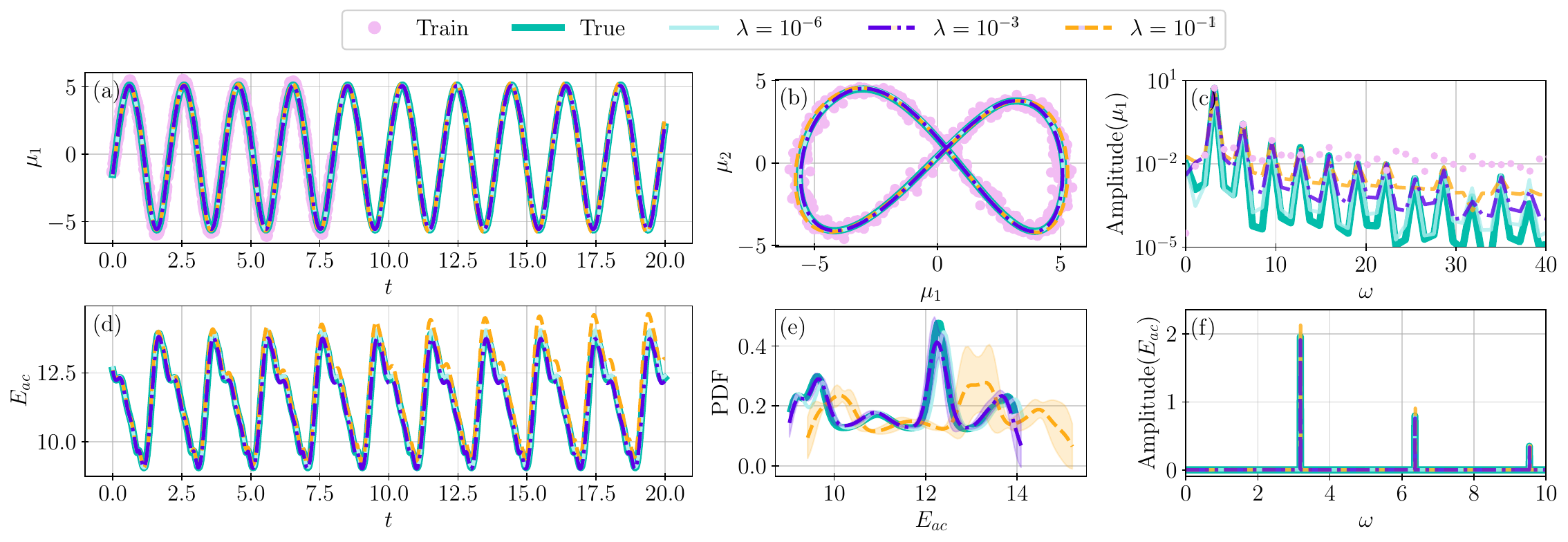}
    \caption{\revise{Thermoacoustic Echo State Network learns the {true limit cycle attractors of the original system} from noisy data. Short-term closed-loop prediction on the training data from $(\beta=2.0,\tau=0.25)$ regime is shown for the first pressure mode as timeseries (a) and for the first and second pressure modes in the phase space (b). The amplitude spectrum (c) is computed for the same time window as the training data (8 time units) for consistency in frequency resolution and shown in logarithmic scale. Short-term prediction (d), probability distribution function (PDF) (e), and amplitude spectrum (f) of the acoustic energy are shown as in Fig.~\ref{fig:lco}}.}\label{fig:noisy_lco}
\end{figure}

\begin{figure}
    \includegraphics[width=\linewidth]{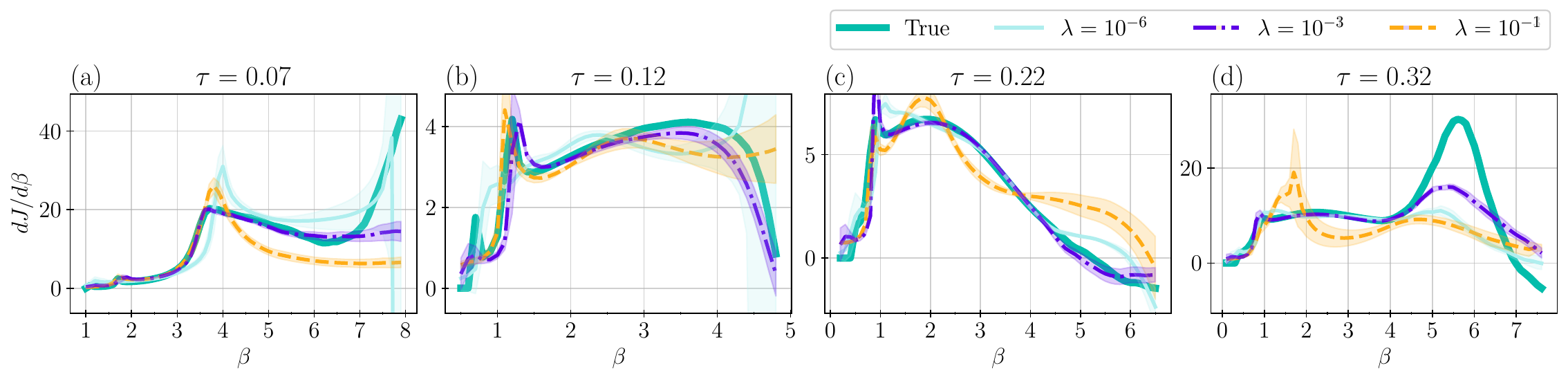}
    \caption{\revise{Regularization with an optimal Tikhonov coefficient $\lambda = 10^{-3}$ improves inference of sensitivity to heat release strength in case of noise level 5\% (See Fig.~\ref{fig:noise_validation}). Shown $d\mathcal{J}/d\beta$ at fixed (a) $\tau = 0.07$, (b) $\tau = 0.12$, (c) $\tau = 0.22$, (d) $\tau =0.32$. The mean of the sensitivity of an ensemble of 5 realisations of the ESN is shown for different regularization coefficients, one standard deviation from the mean is plotted in lighter colour.}}\label{fig:noisy_sensitivity_beta}
\end{figure}

\begin{figure}
    \includegraphics[width=\linewidth]{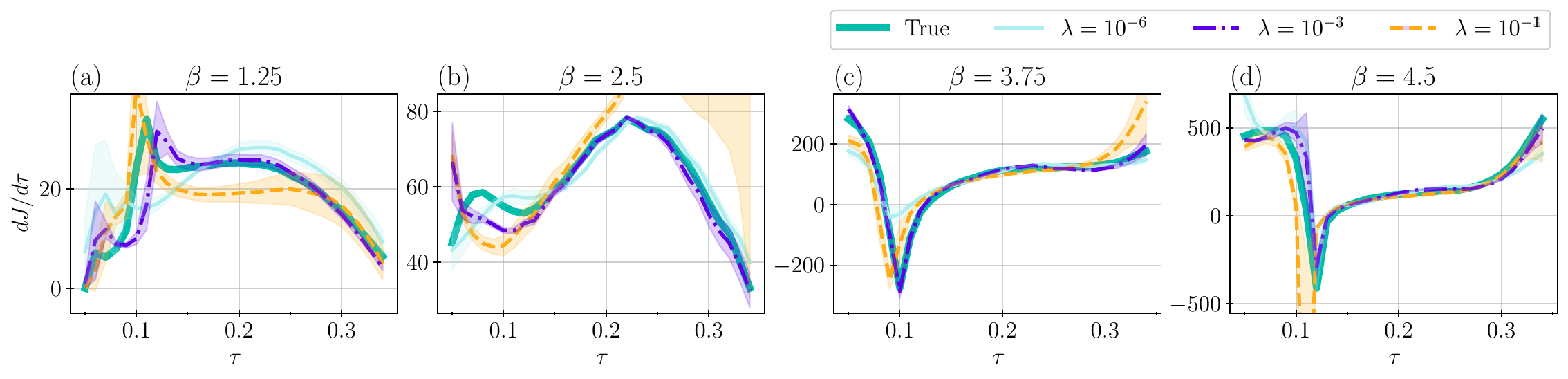}
    \caption{\revise{Regularization with an optimal Tikhonov coefficient $\lambda = 10^{-3}$ improves inference of sensitivity to time delay $\tau$ in case of noise level 5\% (See Fig.~\ref{fig:noise_validation}). Shown $d\mathcal{J}/d\tau$ at fixed (a) $\beta = 1.25$, (b) $\beta = 2.5$, (c) $\beta = 3.75$, (d) $\beta = 4.5$. The mean of the sensitivity of an ensemble of 5 realisations of the ESN is shown for different regularization coefficients, one standard deviation from the mean is plotted in lighter colour.}}\label{fig:noisy_sensitivity_tau}
\end{figure}

For the 5\% noise level case, we show the performance of closed-loop prediction on a parameter regime $(\beta=2.0,\tau=0.25)$, which is in the training set, in Fig.~\ref{fig:noisy_lco}. The T-ESN successfully learns the limit cycle attractor from noisy data and does not become unstable when the training data is noisy. We compute the adjoint sensitivities for the same test cases in Sec.\ref{sec:results_adj}, these are shown in \revise{Figs.~\ref{fig:noisy_sensitivity_beta}~and~\ref{fig:noisy_sensitivity_tau} for the heat release strength $\beta$ and time delay $\tau$, respectively}. The sensitivity computations are affected by the noisy training especially in small $\tau$ regimes ($\tau \leq 0.1$) that are unseen during training, \revise{e.g., Fig.~\ref{fig:noisy_sensitivity_beta}~(a,b)}. These errors are alleviated by increasing the regularization. However, using a very large regularization might decrease the performance, for example in higher $\tau$ regimes ($\tau \geq 0.25$) as shown in Figs.~\ref{fig:noisy_lco}~and~\ref{fig:noisy_sensitivity_tau}(b).

\section{Bifurcations to nonlinear regimes}\label{sec:results_chaos}
\begin{table}
    \caption{Optimal hyperparameters of the thermoacoustic ESN~\eqref{eq:thermoacoustic_esn_step} trained on data from $\beta = \{1.0, 2.0, 3.0, 4.0, 5.0\}$ and $\beta = \{6.0, 6.5, 7.0, 7.5, 8.0\}$ regimes ($\tau = \{0.1, 0.15, 0.2, 0.25, 0.3\}$). Refer to~\eqref{eq:thermoacoustic_esn_step} for the hyperparameters.}
    
    \begin{tabular}{c|c|c|c|c|c|c|c|c|c}
                &  $\rho$ & $\sigma_{in}$ & $\sigma_{f}$ & $\sigma_{\beta}$ & $k_{\beta}$ & $\alpha$ & $\lambda$ \\ \hline
                &  0.01 -- 1.0 & 0.01 -- 2.0 &  0.01 -- 2.0 & 0.01 -- 2.0 & -10.0 -- 10.0 & 0.01 -- 1.0 &  $10^{-6}$ -- $10^{-1}$ \\
        ESN $\beta_{train} = [1,5]$  & 0.0125 & 0.0790 & 1.0501 & 0.1092 & -10.0 & 0.0363 & $10^{-6}$ \\
        ESN $\beta_{train} = [6,8]$   & 0.1602 & 0.0122 & 1.296 & 0.3800 & -4.008 & 0.6369 & $10^{-6}$ \\
    \end{tabular}\label{tab:hyperparameters_2}
\end{table}

\begin{figure}
    \includegraphics[width=\linewidth]{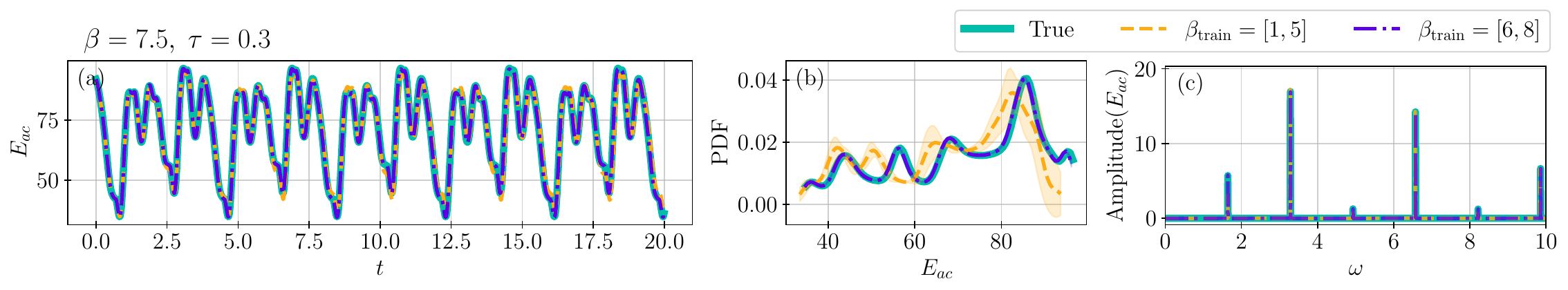}
    \includegraphics[width=\linewidth]{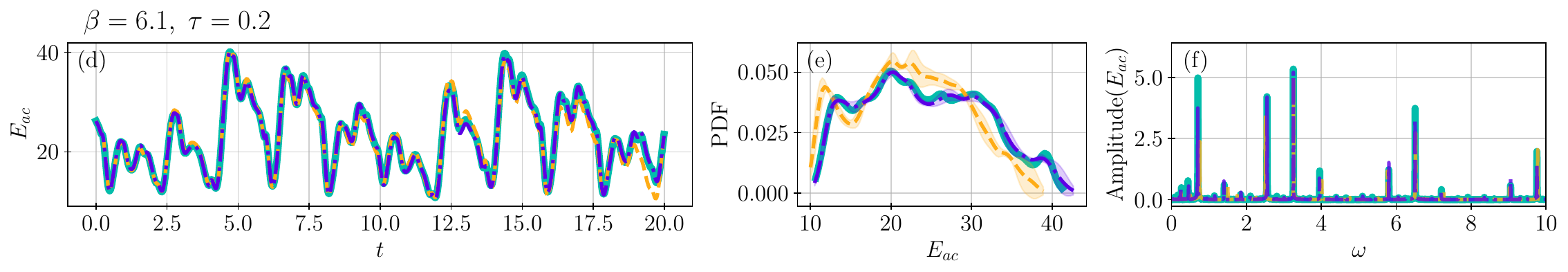} 
    \caption{\revise{Thermoacoustic Echo State Network captures bifurcations to period-doubling and quasi-periodic regimes. Short-term prediction, statistics, and amplitude spectrum of (a-f) an unseen limit cycle regime where period-doubling occurs ($\beta = 7.5, \; \tau = 0.3$), (g-l) an unseen quasi-periodic regime ($\beta = 6.1, \; \tau = 0.2$)}}\label{fig:bifns_periodic}
\end{figure}

\begin{figure}
    \includegraphics[width=\linewidth]{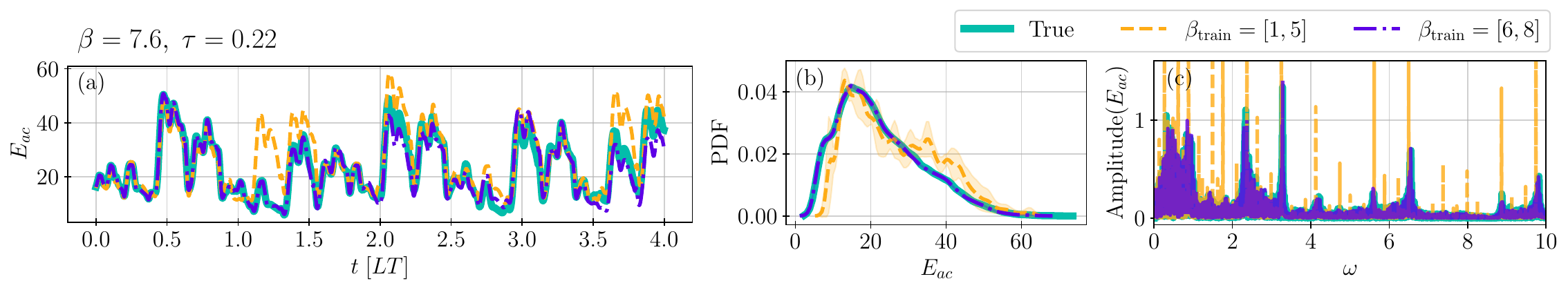}
    \includegraphics[width=\linewidth]{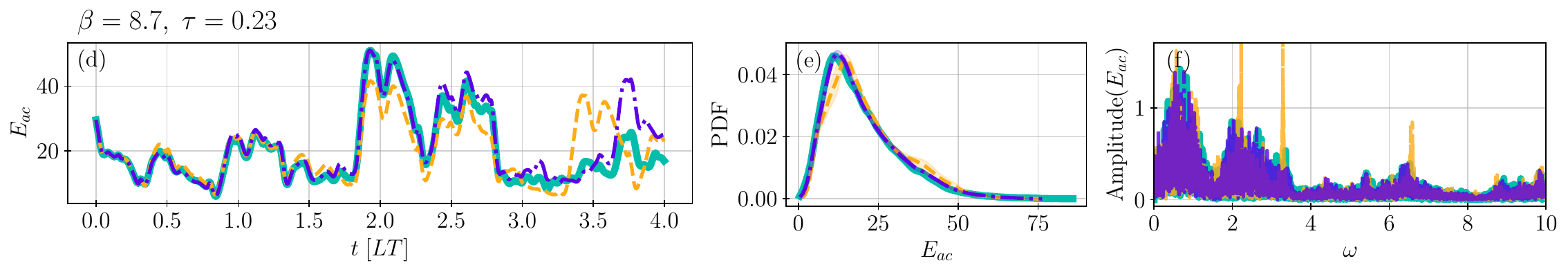} 
    \caption{\revise{Thermoacoustic Echo State Network captures bifurcations to chaotic regimes. Prediction, statistics, and amplitude spectrum of (a-f) an unseen chaotic regime ($\beta = 7.6, \; \tau = 0.22$) with a Lyapunov time of 8.5, and (g-l) an unseen chaotic regime ($\beta = 8.7, \; \tau = 0.23$) with a Lyapunov time of 3.9.}}\label{fig:bifns_chaotic}
\end{figure}

\begin{figure}
    \includegraphics[width=\linewidth]{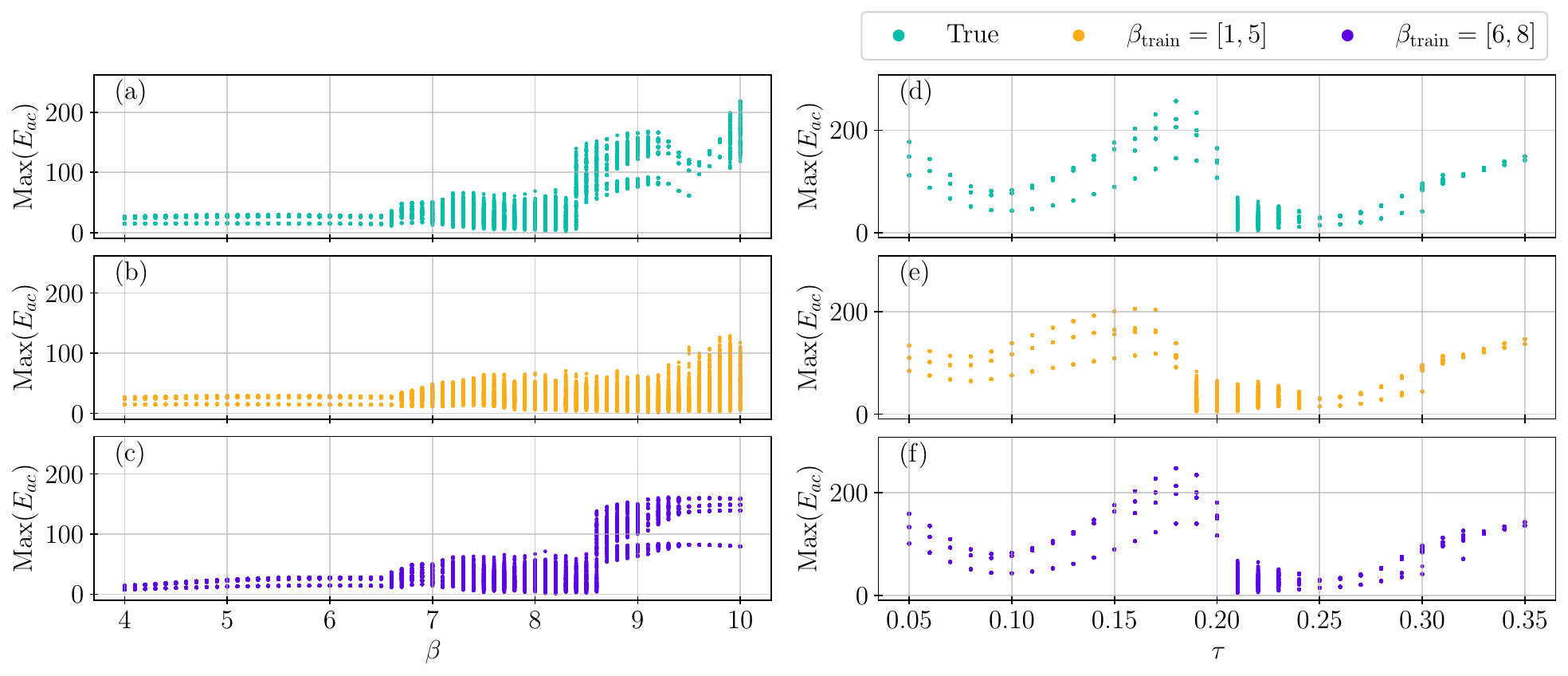}
    \caption{\revise{Bifurcation diagrams obtained from the true system (top row), thermoacoustic Echo State Network trained on $\beta_{train} = [1,5]$ (middle row), and trained on $\beta_{train} = [6,8]$ (bottom row). The peaks of the acoustic energy are plotted. Left half (a,b,c) shows varying $\beta$ for fixed $\tau = 0.22$, and right half (d,e,f) shows varying $\tau$ for fixed $\beta = 7.6$.}}\label{fig:bifn_diagram}
\end{figure}

In Sec.~\ref{sec:results_learn}, we discuss that the ESN predicts bifurcations from limit cycle oscillations to fixed points when the heat release strength $\beta$ is decreased. On the other hand, if $\beta$ is increased, the thermoacoustic system displays bifurcations from limit cycle oscillations to chaos via period-doubling and quasiperiodic routes~\citep{huhn2020StabilitySensitivityOptimisation}. In this section, we investigate whether the ESN can also predict these bifurcations and attractors. We train and validate a separate T-ESN on data generated by larger values $\beta = \{6.0, 6.5, 7.0, 7.5, 8.0\}$ (for brevity, we refer to this dataset as $\beta_{train} = [6.0-8.0]$). The reservoir size and connectivity are 1200 and 20, respectively as before. The optimal hyperparameters found for this case are provided in Table~\ref{tab:hyperparameters_2} along with the hyperparameters for the previous case when the training was performed with data from $\beta = \{1.0, 2.0, 3.0, 4.0, 5.0\}$ (for brevity, we refer to this dataset as $\beta_{train} = [1.0-5.0]$). The faster changing dynamics, e.g., chaotic behaviour, in the larger $\beta$ regimes might be responsible for the increase in the spectral radius and leak rate. 

\revise{We examine the T-ESN's  prediction performance across different nonlinear regimes. First, in Fig.~\ref{fig:bifns_periodic}~(a-c) we show a limit cycle regime in which period-doubling occurs ($\beta = 7.5, \; \tau = 0.3$), and in Fig.~\ref{fig:bifns_periodic}~(d-f) a quasi-periodic regime ($\beta = 6.1, \; \tau = 0.2$). Second, we show two chaotic thermoacoustic regimes with different Lyapunov times; in Fig.~\ref{fig:bifns_chaotic}~(a-f) a regime ($\beta = 7.6, \; \tau = 0.22$) with a Lyapunov time of 8.5 time units and in Fig.~\ref{fig:bifns_chaotic}~(g-l) a regime ($\beta = 8.7, \; \tau = 0.23$) with a Lyapunov time of 3.9 time units. Lyapunov time (LT) is the characteristic timescale, by which two nearby trajectories diverge exponentially in a chaotic system and is an indicator of the predictability of a system. The ESN trained on $\beta_{train} = [1.0-5.0]$ can qualitatively predict the bifurcations and capture the statistics and dominant frequencies correctly, especially for the non-chaotic regimes. Training the T-ESN on $\beta_{train} = [6.0-8.0]$, which encompasses the regimes that we are testing, increases accuracy of both short- and long-term predictions.}

\revise{The ability of the T-ESNs trained on different $\beta$ regimes to predict bifurcations is further shown in Fig.~\ref{fig:bifn_diagram} in the bifurcation diagrams for fixed $\tau = 0.22$ (a-c) and $\beta = 7.6$ (d-f) cases. This ability of the parameter-aware ESN to predict bifurcations to quasiperiodic or chaotic regimes at unseen parameter values is in itself not an obvious result as the network only had access to limit cycle regimes at lower $\beta$ values for training and validation. 
When we increase $\beta$ and a bifurcation occurs, the acoustic energy increases with a discontinuous jump, which indicates that the amplitude of the oscillations abruptly change. Therefore, the prediction of these regimes becomes an extrapolation problem because the range of data we are interested in is different from the range of data the network was trained on. In this case, incorporating physical knowledge via the T-ESN architecture enables extrapolation performance. The bifurcation diagrams confirm that in the higher $\beta$ regimes, the T-ESN trained on $\beta_{train} = [1.0-5.0]$ can qualitatively capture the bifurcations to quasiperiodic and chaotic regimes, however is not as quantitatively accurate as the T-ESN trained on $\beta_{train} = [6.0-8.0]$. (The reverse is true for  lower $\beta$ regimes.) When $\beta$ is further increased beyond $\beta=10.0$, the quantitatively accurate performance of the T-ESN trained on $\beta_{train} = [6.0-8.0]$ also starts diminishing.}

\begin{figure}
    \includegraphics[width=\linewidth]{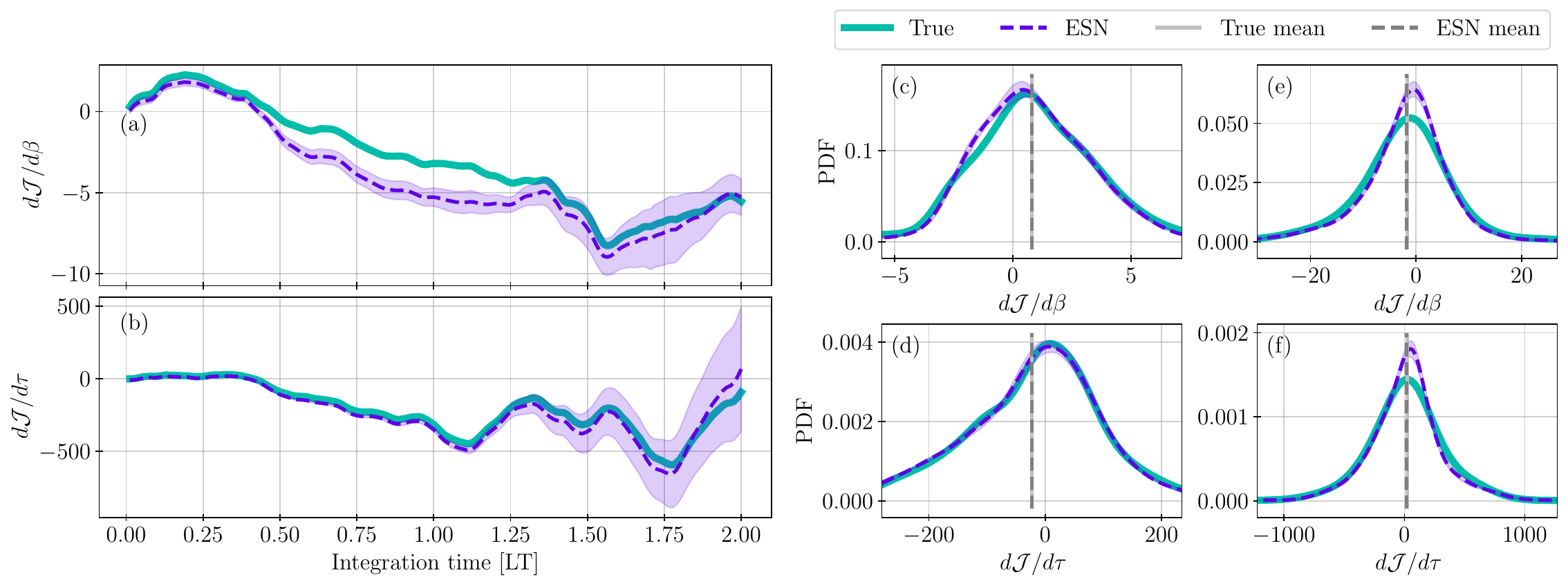}
    \caption{Computation of chaotic adjoint sensitivities using thermoacoustic Echo State Network. The adjoint sensitivities for a single trajectory at $(\beta=7.6, \; \tau = 0.22)$ is shown on the left (a,b). The probability distribution of the sensitivity is computed from 2000 short trajectories with integration time of 0.5 Lyapunov Time (c,d) and 1.0 Lyapunov Time (e,f).}\label{fig:chaotic_sensitivity}
\end{figure}

In chaotic systems, due to the positive Lyapunov exponent, the adjoint system is unstable and the adjoint sensitivities diverge~\citep{lea2000SensitivityAnalysisClimate,huhn2020StabilitySensitivityOptimisation}. \revise{The ESN learns and replicates the underlying chaotic dynamics of the system under investigation~\citep[e.g.,][]{margazoglou2023StabilityAnalysisChaotic}. Because the ESN is a dynamical system~\eqref{eq:parameter_esn_step_closed_loop}, the adjoint sensitivity in chaotic regimes is unstable in the long-term~\citep{huhn2020StabilitySensitivityOptimisation}}. This means that we can only integrate adjoint equations for short time windows. Using the T-ESN trained on $\beta_{train} = [6.0-8.0]$, we compute the adjoint sensitivities of short-time trajectories for the ($\beta = 7.6, \; \tau = 0.22$) regime (shown in Fig.~\ref{fig:bifns_chaotic}~(a-c)). Figs.~\ref{fig:chaotic_sensitivity}~(a,b) show the evolution of the adjoint sensitivities with integration time for one trajectory for an ensemble of 5 T-ESNs. For short integration times, the T-ESN predicts the chaotic sensitivities close to the original system, but diverges as the integration time increases due to the chaotic nature of the system. One approach to estimate the sensitivity of chaotic systems is by taking the mean of an ensemble of trajectories~\citep{lea2000SensitivityAnalysisClimate}. In Figs.~\ref{fig:chaotic_sensitivity}~(c,d)~and~(e,f), we show the probability distributions and the mean of the adjoint sensitivities for 2000 trajectories for integration times of 0.5 LT and 1 LT, respectively. In order to obtain the PDFs, we let the T-ESN run autonomously without any washout data from the original system. For an integration time of 0.5 LT, we can recover the PDF as well as the mean, while for 1.0 LT, low statistical moments are matched. 

\begin{figure}
    \includegraphics[width=\linewidth]{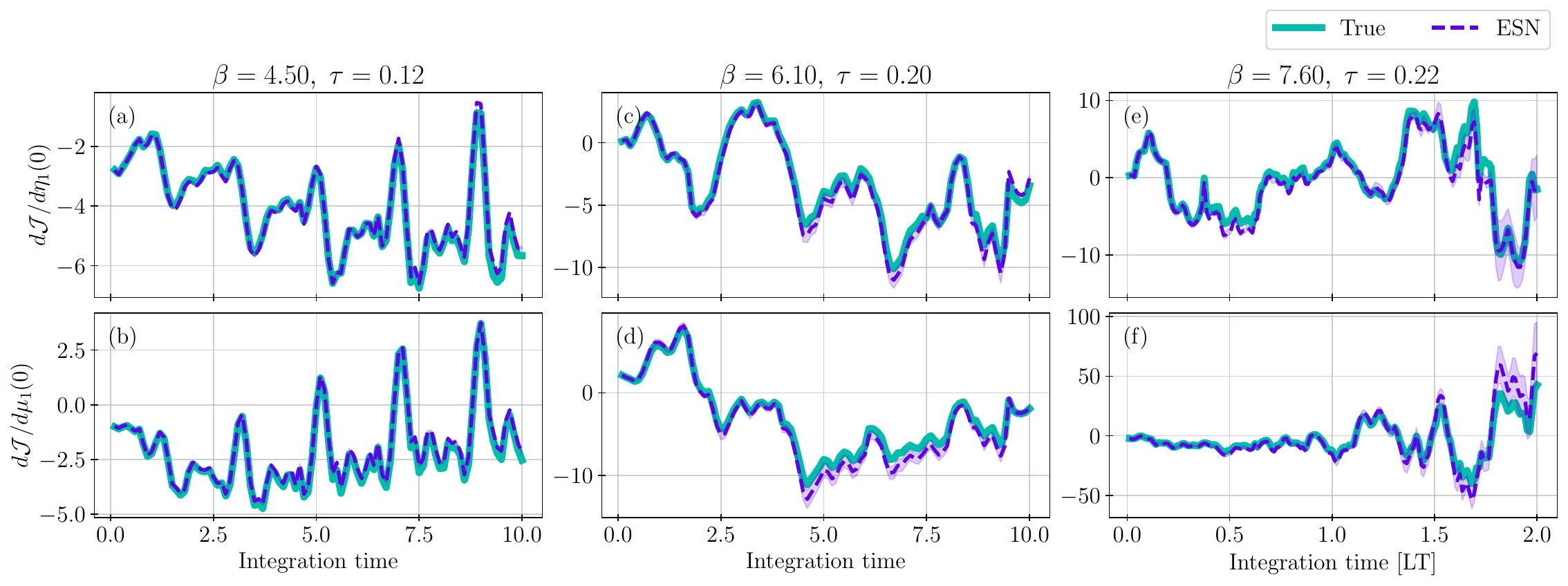}
    \caption{\revise{Computation of adjoint sensitivities to initial conditions using thermoacoustic Echo State Network. The adjoint sensitivity for a single trajectory at unseen (a,b) limit cycle $(\beta=4.5, \; \tau = 0.12)$, (c,d) quasiperiodic $(\beta=6.1, \; \tau = 0.2)$, (e,f) chaotic $(\beta=7.6, \; \tau = 0.22)$ regimes. The sensitivity of the acoustic energy of the final flow state to perturbation to the first velocity mode $\eta_1$ (top row) and to the first pressure mode $\mu_1$ (bottow row) are shown for increasing integration times.}}\label{fig:init_sensitivity}
\end{figure}

\revise{\section{Computation of sensitivities to initial conditions}\label{sec:init_sensitivity}}
\revise{The adjoint method  allows us to efficiently compute sensitivities to initial conditions. In this section, we infer the sensitivity to initial conditions, i.e., $\vector{y}(0) = [\vector{\bm{\eta}}(0); \; \vector{\bm{\mu}}(0)]$, using the T-ESNs trained in the previous sections. The objective functional is the acoustic energy of the final flow state. The sensitivities to perturbations to the first velocity and pressure modes, $\eta_1$ and $\mu_1$, are shown in Fig.~\ref{fig:init_sensitivity} for a trajectory from exemplary limit cycle $(\beta=4.5, \; \tau = 0.12)$, quasiperiodic $(\beta=6.1, \; \tau = 0.2)$, and chaotic $(\beta=7.6, \; \tau = 0.22)$ regimes. To assess the accuracy of the estimates, we repeat this procedure for multiple independent trajectories. Because in gradient-based optimization we are interested in the gradient direction, we compute the angle between the sensitivity gradients of the original system and the T-ESN estimates as an indicator of the accuracy. The mean of the computed angles over 20 trajectories and 5 realisations of T-ESN are provided in Tab.~\ref{tab:angles} for different integration times. We find that T-ESN achieves high accuracy for all integration times in the limit cycle regimes and for short integration times in quasiperiodic and chaotic regimes.}

\begin{table}
    \caption{\revise{Angle between the sensitivity gradients of the original system to initial conditions and the estimates from thermoacoustic Echo State Network (T-ESN). The mean angle over 20 independent trajectories and 5 realisations of T-ESN is given in degrees.}}
    \begin{tabular}{lll}
        \begin{tabular}{c c |c|c}
                &    & Limit cycle & Quasiperiodic \\
                &    & $\; (\beta=4.5, \; \tau = 0.12) \;$ &  $\;(\beta=6.1, \; \tau = 0.2)\;$ \\ \cline{2-4}
                \multirow{3}{*}{\begin{tabular}{c}Integration \\ Time\end{tabular}} & 2.0  & $2.2$  & $6.8$   \\
                                                                  & 4.0  & $2.3$  & $12.8$  \\
                                                                  & 10.0 & $3.2$  & $16.2$ 
        \end{tabular}
        &\quad
        \begin{tabular}{c|c}
                & Chaotic \\
                & $\;(\beta=7.6, \; \tau = 0.22)\;$ \\ \cline{1-2}
        0.5 LT & $8.0$  \\
        1.0 LT & $15.5$ \\
        2.0 LT & $34.3$
        \end{tabular}
    \end{tabular}
    \label{tab:angles}
\end{table}

\revise{\section{Discussion}\label{sec:limit}}
\revise{In this section, we offer a comparison of the proposed ESN-based adjoint solver with traditional, i.e., physics-based, adjoint solvers for gradient-based optimization. 
With the physics-based approach, we need (i) governing equations (plus boundary conditions and initial conditions), (ii) a numerical scheme to solve them, (iii) the linearization (tangent model), and (iv) its dual with respect to a chosen inner product (adjoint code). We need to derive/modify the adjoint code any time that any of the above is modified. On the other hand, in the ESN-based approach the adjoint code is derived once and does not change, i.e., the adjoint equations of the ESN described by Eqs.~\eqref{eq:adjoint}~and~\eqref{eq:adjoint_r0} are independent of the data and system under investigation. Similarly, the adjoint equations of the time-delayed ESN described by Eq.~\ref{eq:adjoint_delayed} are applicable to any system with a constant time-delay. The ESN-based approach is also particularly attractive when we have experimental data, in which the equations might  be unknown (or partly unknown) and the data contains noise, because it learns the dynamics from observables.}\\ 

\revise{As a data-driven approach, the ESN-based method relies on (i) having sufficient data that covers the solution space of the problem, and (ii) training and validation of the ESN. 
When utilizing a parameter-aware ESN for the computation of parameter sensitivities, as the number of system's parameters increase, more training regimes are required. 
This is in contrast with physics-based adjoint solvers, which are derived from the system's equations that generalise over the parameter space. Thus, if the ESN-based adjoint solver is sought to replace a physics-based adjoint solver, the advantage can vary from case to case depending on the cost of data acquisition and simulation time.   
Because the network learns a continuous parameterization of the dynamics even if it is trained on data from a discrete set of parameters, the adjoint can be computed around any set of parameters. (With a continuous parameterization of the objective functional, a future research direction is to combine Bayesian optimisation~\citep[e.g.,][]{huhn2022GradientfreeOptimizationChaotic}, which relies on sampling via the ESN, and adjoint-based ESN gradients.) The physics-based adjoint solver depends on the modelling assumptions, whereas the accuracy of the ESN-based adjoint solver depends on data and the hyperparameters. In high-dimensional systems, the ESN-based adjoint solver can be combined with a data-driven model reduction techniques, e.g., autoencoding as in \citep{racca2023PredictingTurbulentDynamicsa}.}

\revise{Finally, regarding computational costs, the training of the ESN is faster compared to other neural networks that are trained with stochastic gradient descent and backpropagation, and the T-ESN runs faster than the original solver when simulating the thermoacoustic system considered in this paper. Although this can be case dependent, in general the propagation of the ESN equations is fast thanks to the sparse weight matrices even though the reservoir state can be high dimensional (here, 1200). An optimization of the code for matrix multiplications on a GPU, e.g., using JAX~\citep{kochkov2021MachineLearningAccelerated}, is scope for future work.}


\section{Conclusion}\label{sec:conclusion}
Adjoint methods are established tools for gradient-based optimization. Although adjoint methods offer an accurate and efficient gradient computation, they are problem specific. In this work, we propose a strategy to infer adjoint sensitivities from data (observables). We showcase the method on a prototypical nonlinear time-delayed thermoacoustic system. 
This system is described by a nonlinear wave equation. The parameters of interest are the heat release strength and the time-delay at the heat source location. By varying these parameters, the system displays fixed-point, limit-cycle, quasiperiodic, and chaotic solutions. The cost functional is the time-averaged acoustic energy, which we wish to minimize to suppress an instability. 

First, we propose a parameter-aware Echo State Network (ESN). This network learns the parameterized dynamics, i.e., it infers how the dynamics change with varying parameters, thereby making predictions for parameters that are not used during training. Second, we derive the adjoint parameter-aware Echo State Network, which is used to compute the adjoint sensitivities of the physical system. \revise{Although the proposed framework is general and has been applied to prototypical low-dimensional dynamical systems without any modification of the standard architecture, we find that for the thermoacoustic system of interest embedding physical knowledge improves generalizability and requires less data.} Therefore, third, we design a parameter-aware ESN motivated by the physics of thermoacoustics (thermoacoustic ESN, or T-ESN in short) by 
(i) making the ESN explicitly time-delayed, and 
(ii) designing the input matrix in consideration of acoustic physics. We extend the derivation of the adjoint to the time-delayed case. We find that the T-ESN accurately predicts the dynamics of different nonlinear regimes, and successfully infers the adjoint sensitivities. These results are validated against the ground truth generated from integration of the direct and adjoint equations of the original system. We optimize the acoustic energy with steepest gradient-descent, in which the gradients are computed by only one T-ESN. Once the T-ESN has been trained, the algorithm runs  autonomously, i.e., we do not run any additional simulations of the original model. As the optimization proceeds with the iterations, the acoustic energy and the amplitude of the oscillations decrease, finally reaching a fixed-point solution with zero acoustic energy. Fourth, we investigate the robustness to noisy data and chaotic regimes. The T-ESN can qualitatively predict bifurcations to nonlinear regimes beyond the training range of parameter values. Quantitative accuracy increases when the range of training parameters covers the regime in which we are interested. \revise{Finally, in addition to parameter sensitivities, we infer sensitivities to initial conditions.}
This work opens up possibilities for data-driven \revise{gradient-based} optimization without developing code-specific adjoint solvers.
\\
\par The code is available at \citep{github_repo}.

\begin{acknowledgments}
{This research has received financial support from the ERC Starting Grant No. PhyCo 949388 and UKRI AI for Net Zero grant EP/Y005619/1. L.M. is also grateful for the support from the grant EU-PNRR YoungResearcher TWIN ERC-PI\_0000005.}
\end{acknowledgments}

\appendix
\section{Adjoints of discrete-time dynamical systems with and without time delay}\label{sec:appendix_adjoint}
\subsection{Adjoint of a discrete-time system}
The following derivation of the adjoint sensitivity applies to any discrete map of choice that can be expressed as
\begin{equation}\label{eq:discrete_map}
    \vector{x}(i) = \vector{f}(\vector{x}(i-1), \vector{p}),
\end{equation}
where $\vector{x}(i) \in \mathbb{R}^{n_x}$ is the state vector at time step $i$.
Equation~\eqref{eq:discrete_map} is rewritten as a constraint
\begin{equation}\label{eq:constraint}
    \vector{F}(\vector{x}(i),\vector{x}(i-1),\vector{p}) \coloneqq \vector{x}(i) - \vector{f}(\vector{x}(i-1), \vector{p}) = 0.
\end{equation}
We consider a time-averaged quantity as the objective functional
\begin{equation}\label{eq:objective}
    \mathcal{J}(\vector{x},\vector{p}) \coloneqq \frac{1}{N}\sum_{i = 1}^N \tilde{\mathcal{J}}(\vector{x}(i), \vector{p}).
\end{equation}
Following~\citep{magri2019AdjointMethodsDesign}, the problem of minimizing the objective functional~\eqref{eq:objective} under the constraints~\eqref{eq:constraint} can be formulated as a Lagrangian optimization problem
\begin{equation}\label{eq:lagrangian}
    \mathcal{L} \coloneqq \mathcal{J}(\vector{x},\vector{p}) - \langle\vector{q}^+, \vector{F}\rangle,
\end{equation}
where $\vector{q}^+$ are the Lagrange multipliers. The inner product $\langle\vector{q}^+, \vector{F}\rangle$ is given by
\begin{equation}
    \langle\vector{q}^+, \vector{F}\rangle \coloneqq \sum_{i = 1}^N \vector{q}^{+\top}(i) \vector{F}(i),
\end{equation}
where we denote $\vector{F}(i) = \vector{F}(\vector{x}(i),\vector{x}(i-1),\vector{p})$ for brevity.
Because $d\mathcal{J}/d\vector{p} = d\mathcal{L}/d\vector{p}$, we seek the gradient $d\mathcal{L}/d\vector{p}$
\begin{equation}
    \frac{d\mathcal{L}}{d\vector{p}} = \frac{1}{N}\sum_{i = 1}^N \frac{\partial\tilde{\mathcal{J}}}{\partial\vector{p}} + \frac{\partial\tilde{\mathcal{J}}}{\partial\vector{x}(i)}\frac{d\vector{x}(i)}{d\vector{p}} 
    - \sum_{i = 1}^N 
    \vector{q}^{+\top}(i)
    \left(
        \frac{\partial \vector{F}(i)}{\partial \vector{p}}
        + \frac{\partial \vector{F}(i)}{\partial \vector{x}(i)}\frac{d\vector{x}(i)}{d\vector{p}}
        + \frac{\partial \vector{F}(i)}{\partial \vector{x}(i-1)}\frac{d\vector{x}(i-1)}{d\vector{p}}
    \right). 
    \end{equation}
Without loss of generality, we assume that the objective function does not explicitly depend on $\vector{p}$ and therefore $\partial\tilde{\mathcal{J}}/\partial\vector{p}$ is zero. Writing out the terms of the summation, we have
\begin{equation}
    \begin{split}
        \frac{d\mathcal{L}}{d\vector{p}} =  \sum_{i = 1}^N & -\vector{q}^{+\top}(i)\frac{\partial \vector{F}(i)}{\partial \vector{p}}  \\
        & + \frac{1}{N}\frac{\partial\tilde{\mathcal{J}}(\vector{x}(1))}{\partial\vector{x}(1)}\frac{d\vector{x}(1)}{d\vector{p}} - \vector{q}^{+\top}(1)\frac{\partial \vector{F}(1)}{\partial \vector{x}(1)}\frac{d\vector{x}(1)}{d\vector{p}} - \vector{q}^{+\top}(1)\frac{\partial \vector{F}(1)}{\partial \vector{x}(0)}\frac{d\vector{x}(0)}{d\vector{p}} \\
        & + \frac{1}{N}\frac{\partial\tilde{\mathcal{J}}(\vector{x}(2))}{\partial\vector{x}(2)}\frac{d\vector{x}(2)}{d\vector{p}} - \vector{q}^{+\top}(2) \frac{\partial \vector{F}(2)}{\partial \vector{x}(2)}\frac{d\vector{x}(2)}{d\vector{p}}- \vector{q}^{+\top}(2)\frac{\partial \vector{F}(2)}{\partial \vector{x}(1)}\frac{d\vector{x}(1)}{d\vector{p}} \\
        & \vdots \\
        & + \frac{1}{N}\frac{\partial\tilde{\mathcal{J}}(\vector{x}(N))}{\partial\vector{x}(N)}\frac{d\vector{x}(N)}{d\vector{p}} - \vector{q}^{+\top}(N)\frac{\partial \vector{F}(N)}{\partial \vector{x}(N)}\frac{d\vector{x}(N)}{d\vector{p}} - \vector{q}^{+\top}(N)\frac{\partial \vector{F}(N)}{\partial \vector{x}(N-1)}\frac{d\vector{x}(N-1)}{d\vector{p}}.
    \end{split}
\end{equation}
Because of Eq.~\eqref{eq:discrete_map}, the terms $\partial \vector{F}(i)/\partial \vector{x}(i)$ are equal to the identity matrix, and $d\vector{x}(0)/d\vector{p}$ is equal to zero. We choose the Lagrange multipliers to eliminate the terms $d\vector{x}(i)/d\vector{p}$, the size of which grow with the dimension of $\vector{p}$. For this purpose, we gather the terms multiplying $d\vector{x}(i)/d\vector{p}$, thereby obtaining the evolution equations~\eqref{eq:adjoint} for the Lagrange multipliers, i.e., the adjoint variables,
\begin{equation}
    \begin{split}
    & \frac{1}{N}\frac{\partial\tilde{\mathcal{J}}(\vector{x}(1))}{\partial\vector{x}(1)} - \vector{q}^{+\top}(1) - \vector{q}^{+\top}(2)\frac{\partial \vector{F}(2)}{\partial \vector{x}(1)} = 0\\
    & \vdots \\
    & \frac{1}{N}\frac{\partial\tilde{\mathcal{J}}(\vector{x}(N))}{\partial\vector{x}(N)} - \vector{q}^{+\top}(N) = 0,
    \end{split}
\end{equation}
where from Eq.~\eqref{eq:discrete_map}, $\partial \vector{F}(i)/\partial \vector{x} (i-1) = -\partial \vector{f}(i)/\partial \vector{x} (i-1)$.

\subsection{Adjoint of a time-delayed discrete-time system}
The discrete map for the time-delayed can be expressed as
\begin{equation}\label{eq:constraint_delayed}
    \vector{F}(\vector{x}(i),\vector{x}(i-1),\vector{x}(i-1-N_{\tau}),\vector{p}) = 0,
\end{equation}
where $N_{\tau}$ is the time-delay in discrete time steps. The Lagrangian~$\mathcal{L}$ is formulated similarly to Eq.~\eqref{eq:lagrangian}. The gradient $d\mathcal{L}/d\vector{p}$ is given by
\begin{equation}
    \frac{d\mathcal{L}}{d\vector{p}} =  \frac{1}{N}\sum_{i = 1}^N \frac{\partial\tilde{\mathcal{J}}}{\partial\vector{p}} + \frac{\partial\tilde{\mathcal{J}}}{\partial\vector{x}(i)}\frac{d\vector{x}(i)}{d\vector{p}} 
    - \sum_{i = 1}^N \vector{q}^{+\top}(i)
    \left(
        \frac{\partial \vector{F}(i)}{\partial \vector{p}}
        + \frac{\partial \vector{F}(i)}{\partial \vector{x}(i)}\frac{d\vector{x}(i)}{d\vector{p}}
        + \frac{\partial \vector{F}(i)}{\partial \vector{x}(i-1)}\frac{d\vector{x}(i-1)}{d\vector{p}}
        + \frac{\partial \vector{F}(i)}{\partial \vector{x}(i-1-N_{\tau})}\frac{d\vector{x}(i-1-N_{\tau})}{d\vector{p}}
    \right).
\end{equation}
Writing out the terms of the summation, we have
\begin{equation}
    \begin{split}
        \frac{d\mathcal{L}}{dp} =  \sum_{i = 1}^N & -\vector{q}^{+\top}(i)\frac{\partial \vector{F}(i)}{\partial \vector{p}}  \\
        & + \frac{1}{N}\frac{\partial\tilde{\mathcal{J}}(\vector{x}(1))}{\partial\vector{x}(1)}\frac{d\vector{x}(1)}{d\vector{p}} - \vector{q}^{+\top}(1)\frac{\partial \vector{F}(1)}{\partial \vector{x}(1)}\frac{d\vector{x}(1)}{d\vector{p}} - \vector{q}^{+\top}(1)\frac{\partial \vector{F}(1)}{\partial \vector{x}(0)}\frac{d\vector{x}(0)}{d\vector{p}} - \vector{q}^{+\top}(1)\frac{\partial \vector{F}(1)}{\partial \vector{x}(-N_{\tau})}\frac{d\vector{x}(-N_{\tau})}{d\vector{p}}\\
        & + \frac{1}{N}\frac{\partial\tilde{\mathcal{J}}(\vector{x}(2))}{\partial\vector{x}(2)}\frac{d\vector{x}(2)}{d\vector{p}} - \vector{q}^{+\top}(2)\frac{\partial \vector{F}(2)}{\partial \vector{x}(2)}\frac{d\vector{x}(2)}{d\vector{p}} - \vector{q}^{+\top}(2)\frac{\partial \vector{F}(2)}{\partial \vector{x}(1)}\frac{d\vector{x}(1)}{d\vector{p}} - \vector{q}^{+\top}(2)\frac{\partial \vector{F}(2)}{\partial \vector{x}(1-N_{\tau})}\frac{d\vector{x}(1-N_{\tau})}{d\vector{p}} \\
        & \vdots \\
        & + \frac{1}{N}\frac{\partial\tilde{\mathcal{J}}(\vector{x}(2+N_{\tau}))}{\partial\vector{x}(2+N_{\tau})}\frac{d\vector{x}(2+N_{\tau})}{d\vector{p}} - \vector{q}^{+\top}(2+N_{\tau})\frac{\partial \vector{F}(2+N_{\tau})}{\partial \vector{x}(2+N_{\tau})}\frac{d\vector{x}(2+N_{\tau})}{d\vector{p}} - \vector{q}^{+\top}(2+N_{\tau})\frac{\partial \vector{F}(2+N_{\tau})}{\partial \vector{x}(1+N_{\tau})}\frac{d\vector{x}(1+N_{\tau})}{d\vector{p}}\\ & - \vector{q}^{+\top}(2+N_{\tau})\frac{\partial \vector{F}(2+N_{\tau})}{\partial \vector{x}(1)}\frac{d\vector{x}(1)}{d\vector{p}} \\
        & \vdots \\
        & + \frac{1}{N}\frac{\partial\tilde{\mathcal{J}}(\vector{x}(N))}{\partial\vector{x}(N)}\frac{d\vector{x}(N)}{d\vector{p}} - \vector{q}^{+\top}(N)\frac{\partial \vector{F}(N)}{\partial \vector{x}(N)}\frac{d\vector{x}(N)}{d\vector{p}} - \vector{q}^{+\top}(N)\frac{\partial \vector{F}(N)}{\partial \vector{x}(N-1)}\frac{d\vector{x}(N-1)}{d\vector{p}} \\ 
        & - \vector{q}^{+\top}(N)\frac{\partial \vector{F}(N)}{\partial \vector{x}(N-1-N_{\tau})}\frac{d\vector{x}(N-1-N_{\tau})}{d\vector{p}}.
    \end{split}
\end{equation}
Because of Eq.~\eqref{eq:discrete_map}, the terms $\partial \vector{F}(i)/\partial \vector{x}(i)$ are equal to the identity matrix, and due to the initial conditions $d\vector{x}(i \leq 0)/d\vector{p}$ are equal to zero. Eliminating $d\vector{x}/d\vector{p}$ terms, the evolution of the adjoint variables is given by
\begin{equation}
    \begin{split}
    & \frac{1}{N}\frac{\partial\tilde{\mathcal{J}}(\vector{x}(1))}{\partial\vector{x}(1)} - \vector{q}^{+\top}(1) - \vector{q}^{+\top}(2)\frac{\partial \vector{F}(2)}{\partial \vector{x}(1)} - \vector{q}^{+\top}(2+N_{\tau})\frac{\partial \vector{F}(2+N_{\tau})}{\partial \vector{x}(1)} = 0,\\
    & \vdots \\
    & \frac{1}{N}\frac{\partial\tilde{\mathcal{J}}(\vector{x}(N))}{\partial\vector{x}(N)} - \vector{q}^{+\top}(N) = 0.
    \end{split}
\end{equation}
We set $\vector{q}^+(i > N) = 0$ because there are no terms with $d\vector{x}(i>N)/d\vector{p}$.

\section{Gradients of the Echo State Network}\label{sec:appendix_grads}
The gradients associated with the ESN are provided here. We first define,
\begin{equation}
    \tilde{\vector{r}}(i) = \frac{\vector{r}(i+1)-(1-\alpha)\vector{r}(i)}{\alpha},
\end{equation}
which is used to compute the derivative of the $\tanh$ term in the ESN equations, e.g.,~\eqref{eq:esn_step}. The Jacobian and the partial derivative with respect to the system's parameters are 
\begin{align}
    \frac{\partial \vector{r}(i+1)}{\partial \vector{r}(i)} & = (1-\alpha)\matrix{I}_{n_r \times n_r} + \alpha\mathrm{diag}(1-\tilde{\vector{r}}^2(i))(\matrix{W}_{in}^{y}\matrix{W}_{out}+\matrix{W}), \\
    \frac{\partial \vector{r}(i+1)}{\partial \vector{p}} & = \alpha \mathrm{diag}(1-\tilde{\vector{r}}^2(i)) \matrix{W}_{in}^{p}\mathrm{diag}(\vector{\bm{\sigma}}_p), 
\end{align}
where $\matrix{I}$ denotes the identity matrix, and $\mathrm{diag}(\cdot)$ denotes a diagonal matrix that has $(\cdot)$ as its diagonal.
For the thermoacoustic ESN, we have the following gradients
\begin{align}
    \frac{\partial \vector{r}(i+1)}{\partial u_f(i - N_{\tau})} & = \alpha \mathrm{diag}(1-\tilde{\vector{r}}^2(i)) \matrix{w}_f \\ 
    \frac{\partial \vector{r}(i+1)}{\partial \vector{r}(i-N_{\tau})} & = \frac{\partial \vector{r}(i+1)}{\partial u_f(i - N_{\tau})} \Psi \matrix{W}_{out},
\end{align}
where $\Psi = [\cos(\pi x_f); \; \cos(2\pi x_f); \;\dots ; \; \cos(n_g \pi x_f) ; \; \vector{0}_{n_g}]^\top$ contains the Galerkin modes associated with velocity and masks the Galerkin modes associated with pressure.

\bibliography{bibliography}
\end{document}